\documentclass[preprint,12pt]{elsarticle}

\usepackage{amssymb}
\usepackage{color}
\usepackage{amsmath}
\usepackage{hyperref}

\newcommand{\diff}[2][]{
  d  #2\mspace{-8mu}^{^{{\tiny #1}}} 
}

\journal{Computer Methods in Applied Mechanics and Engineering}

\begin{document}

\begin{frontmatter}



\title{Optimally accurate operators for partial differential equations}


\author[label1,label2]{Nobuaki Fuji} 
\author[label3,label4]{Thibault Duretz} 

\affiliation[label1]{organization={Université Paris Cité, Institut de physique du globe de Paris, CNRS},
            city={Paris},
            postcode={F-75005}, 
            country={France}}

\affiliation[label2]{organization={Institut universitaire de France},
            city={Paris},
            country={France}}
\affiliation[label3]{organization={Institut für Geowissenschaften, Goethe-Universität Frankfurt},
            city={Frankfurt},
            country={Germany}}
\affiliation[label4]{organization={CNRS, Géosciences Rennes UMR 6118, University Rennes},
            city={Rennes},
            country={France}}

\begin{abstract}
In this contribution, we generalize the concept of \textit{optimally accurate operators} proposed and used in a series of studies on the simulation of seismic wave propagation, particularly based on Geller \& Takeuchi (1995).
Although these operators have been mathematically and numerically proven to be more accurate than conventional methods, the theory was specifically developed for the equations of motion in linear elastic continuous media. Furthermore, the original theory requires compensation for errors from each term due to truncation at low orders during the error estimation, which has limited its application to other types of physics described by partial differential equations.

Here, we present a new method that can automatically derive numerical operators for arbitrary partial differential equations. These operators, which involve a small number of nodes in time and space (compact operators), are more accurate than conventional ones and do not require meshing. Our method evaluates the weak formulation of the equations of motion, developed with the aid of Taylor expansions.

We establish the link between our new method and the classic optimally accurate operators, showing that they produce identical coefficients in homogeneous media. Finally, we perform a benchmark test for the 1D Poisson problem across various heterogeneous media. The benchmarks demonstrate the superiority of our method compared to conventional operators, even when using a set of linear B-spline test functions (three-point hat functions). However, the convergence rate can depend on the wavelength of the material property: when the material property has the same wavelength as that of the field, the convergence rate is O(4), whereas it can be less efficient O(2) for other models.
\end{abstract}

\begin{graphicalabstract}
\includegraphics{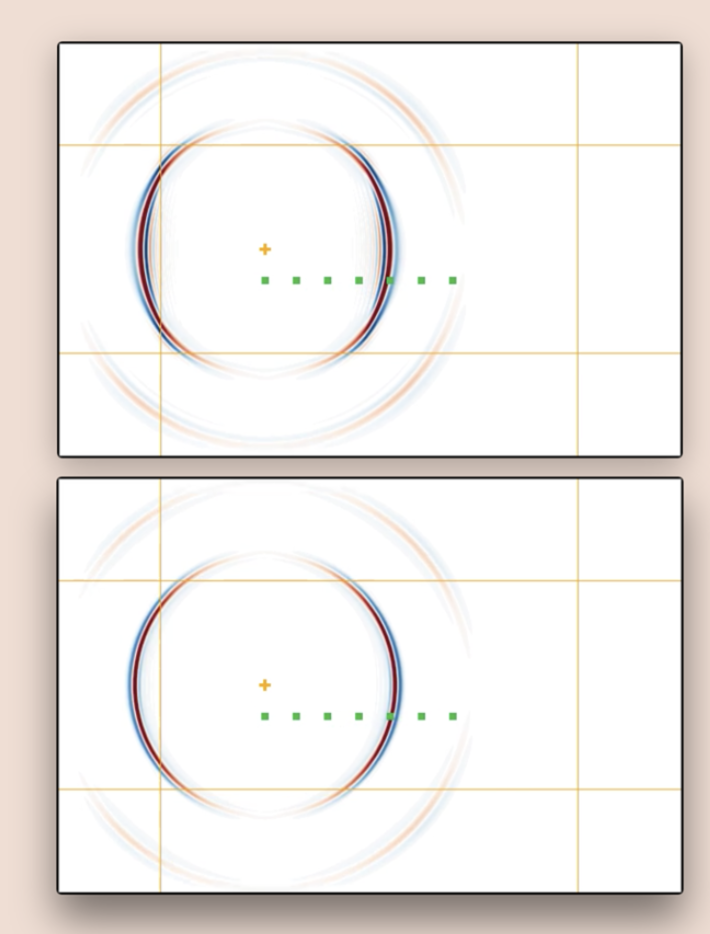}
\end{graphicalabstract}

\begin{highlights}
\item Optimally accurate operators (compact operators) can be obtained automatically for an arbitrary partial differential equation
\item With a compact hat function as a trial function (3 points), optimally accurate operators show a significant improvement in accuracy with respect to the 3-point and 4-point finite-difference method,
 but the convergence rate can depend on the nature of input models.
\end{highlights}

\begin{keyword}
    optimally accurate operators \sep compact  \sep numerical errors \sep seismology \sep geodynamics
\end{keyword}

\end{frontmatter}




\section{Introduction}
In this contribution, we propose a novel methodology that can generate the \textit{optimally accurate operators} for arbitrary partial differential equations. The aim of this study is to propose a general framework for automatically generating numerical operators for a given system of partial differential equations, achieving optimal accuracy while keeping the scheme as simple and computationally efficient as possible. Although there have been numerous efforts to suppress numerical errors while minimizing the computational costs in all the scientific communities, some methods are designed for a specific equation and a specific configuration, with optimizations deeply embedded in the logic (and hence the code), making them difficult to apply to other physical systems or setups. 
The \textit{optimally accurate operator} is one such numerical method, and here we develop a new theory to reinterpret the previously obtained operators from a different perspective, making the method more broadly applicable. 

The optimally accurate operator has been proposed and used in the series of studies on the simulation of seismic wave propagation \cite[e.g.][]{Geller_1995,Geller_1998,Takeuchi_2000,Geller_2011,Hasegawa_2016,Hasegawa_2018}.
During the development of the direct solution method (DSM) for seismic wavefields \cite{Geller_1994}, the group found 
systematic error peaks of their operators for \mbox{1D} elastic wave equation in frequency domain \cite{Geller_1995}.
The direct solution method is a finite-element Galerkin method with a compact support of linear B-spline test functions; but it had encountered this problem of large errors, which happened only in the vicinity of the normal mode (eigenfrequency) of the system.
This is how the optimal accuracy theory was proposed \cite{Geller_1995} (hereafter cited as GT95) based on the formal normal-mode basis. They explained that the numerical errors projected on the normal-mode basis revealed that the numerical error in mass matrix can be canceled by that of the stiffness matrix if the two terms are constructed carefully, i.e. each term of operators does not need to be accurate.
By `modifying' some coefficients of the Galerkin method based on this theory, they were able to acquire a desired numerical errors. However, the concrete detail of derivation of these coefficients has never been demonstrated. After GT95, the direct solution method with optimally accurate operators is extended for \mbox{1D} spherically homogeneous anisotropic anelastic planet-like medium \cite[e.g.][]{Cummins_Computation_1997,Kawai_Complete_2006} and \mbox{3D} Fréchet kernel computation for \mbox{1D} Earth models \cite[e.g.][]{Fuji_2012}, which has been the essence for several deep Earth/Mars exploration studies using waveform inversion \cite[e.g.][]{Fuji_2010,Konishi_2014,Konishi_2020,Fuji_2020,Franken_2020,Jacob_2022}. 

At the same time, the same group developed an optimally accurate operator for the finite-difference scheme applied to the 1D homogeneous elastic equation of motion \cite{Geller_1998}, which was later extended to 2D and 3D media \cite{Geller_1998,Takeuchi_2000} using a predictor-corrector scheme \cite{Geller_2011,Hasegawa_2016}. Since the spectral element method is one of the most popular methods in seismology \cite[e.g.,][]{Komatisch_1998,Komatitsch_2002,NissenMeyer_2014}, the author of this article also developed an optimally accurate operator for the spectral element method \cite[e.g.,][]{Hasegawa_2018}. However, due to its abstract and complicated theory, and due to its lack of simple derivation of operators, the optimally accurate operator has not been a popular numerical method although the community has been aware of its potential \cite[e.g.][]{Moczo_2007,Igel_2016}. 

In this contribution, we generalize the theory without relying on any normal-mode-based methods. In contrast to previous studies, we discovered that the numerical errors of each term do not cancel out; rather, each term can be optimized using a unique approach. Specifically, the method is no longer a Galerkin method but has the following characteristics: (i) it is a finite-element method, as we evaluate the volume integral of the equation of motion multiplied by a set of test functions; and (ii) it has aspects of a finite-difference method, as we compute the partial derivatives of the fields and material coefficients using a Taylor expansion. Due to its finite-element characteristics, it is natural to account for boundary conditions, although this is not the primary focus of the paper. However, because of its finite-difference nature, the method is ready for use once we provide a set of compact coefficients to multiply with the field and material coefficients, which are evaluated at regular nodes - meaning that no specific meshing is required.

This is why the optimally accurate operator bears similarities to other compact finite-difference methods \cite[e.g.][]{Lele1992Nov,Spotz1995Oct,Spotz1996Mar,Carey1997Jul}. These studies do not formally evaluate the weak form of partial differential equations, but by matching the coefficients for partial derivatives in a staggered manner, they ensure the continuity of the functions. For physics involving heterogeneous material properties, a compact scheme for the Poisson equation has been proposed \cite[e.g.][]{Abide2020Dec}.

Recently in seismology, a new methodology called the `Distributional Finite-Difference Method' has been proposed \cite[e.g.][]{Masson_2022,Masson_2023}. We will discuss this methodology in the following sections, but in short, it is a finite-difference method with a strong finite-element character. This method computes the piecewise integral of the field using high-order basis functions. Due to the alternating partial derivatives of the basis functions, the method can achieve high precision for anti-symmetric partial derivatives (i.e., odd-numbered order derivatives).

In this contribution, we focus on the theory behind our novel method, showing its similarity to the classical optimally accurate operator. We also present a numerical example for the 1D Poisson equation. Although the theory is straightforward, its generality may make it less intuitive at first glance. For this reason, we provide a detailed appendix (with different notations) to help guide the reader through the logic more intuitively.

\section{Theory}
In this section, we describe the general framework of optimally accurate operators for partial differential equations. We first introduce the notation and terminology, then we describe the discretisation and the Taylor expansion of the continuous field and material coefficients. We then introduce the B-spline basis functions, which are used to define the optimally accurate operators. Finally, we present the optimally accurate operators for partial differential equations. 
In the following sections, we show the applications of the optimally accurate operators for the wave equation and the heat equation, with more concrete expressions. 
The reader can skip to the examples if the generalised expression below is too abstract.
\subsection{Preparatory notation and terminology}
In this section we define the notation and terminology used in the following sections. We consider a linear or linearised partial differential equation of a strong formulation:
\begin{equation}
    \mathcal{L}_{ji} u_i (\mathbf{y}) - \mathcal{F}_{jk} f_k (\mathbf{y}) =0 \ \ \mathrm{for} \ \ (i,j) \in \{ 1,2,\cdots,M \} \times  \{ 1,2,\cdots,M \}, \ \ \mathbf{y} \in \Omega \subseteq \mathbb{R}^N,
    \label{eq:strongForm}
\end{equation}
and boundary/initial conditions:
\begin{equation}
\mathcal{B}_{ji} u_i (\mathbf{y}) = 0 \ \ \mathrm{for}\ \ (i,j) \in \{ 1,2,\cdots,M \} \times \{ 1,2,\cdots,M \}, \ \ \mathbf{y} \in \partial \Omega \subseteq \mathbb{R}^{N-1},
\label{eq:boundaryConditionStrongForm}
\end{equation}
with $u_i \in U$ the $i$-th component of unknown field with $U$ a real number or complex number space, $\mathcal{L}_{ji}$ and $\mathcal{B}_{ji}$ partial differential operators, with the $j$-th component of excitation, with a partial differential operator $\mathcal{F}_{jk}$ applied to a certain source term $f_k$. $M$ is the dimension of $u_i$, which are the independent components of the field.
While there is no constrains on the form of $\mathcal{F}$, $\mathcal{L}$ should be a square-matrix $M\times M$operator, if the system requires a single unique solution.
$u_i$ and $f_k$ are function of position $\mathbf{y}$ defined on the $N$-dimensional space-time domain  $\Omega$, i.e. $N$ is usually less than or equal to four, (or some other domains such as Fourier domains or spherical harmonic basis domains) and hence $\mathbf{y} \in \Omega \subseteq \mathbb{R}^N$. 

\subsubsection{Locally Cartesian coordinates}
Throughout this paper we assume that $\mathbf{y}$ is locally Cartesian, in the domain related to space and/or time, (as is the case for spherical harmonics for example):
\begin{equation}
\frac{\partial}{\partial y_n} \mathbf{y} = \mathbf{e}_n \ \ \ \textrm{for} \ \ \ n \in \{ 1,2,\cdots,N \}, \forall \mathbf{y} \in \Omega,
\label{eq:localCartesian}
\end{equation}
with $\mathbf{e}_{n}$ a unit vector in $n$-th direction.
The components $i$ and $k$ can be the different types of measurements, such as three components of ground motion and six components of double couple sources in seismology, three-dimensional compornents of particle velocity and pressure field in fluid dynamics (or their dimension-reduced versions), respectively. The Einstein summation convention for repeated subscripts is implicitely used throughout this paper.

\subsubsection{Partial derivatives and antiderivatives}
In order to simplify the expression of partial derivatives of any function $\psi$ defined on $\Omega$, we define:
\begin{equation}
\psi^{(\mathbf{n})}(\mathbf{y})= \psi^{(n_1,n_2,\cdots,n_N)}(\mathbf{y}) \triangleq \partial_{\mathbf{y}}^{\mathbf{n}} \psi(\mathbf{y})= \frac{\partial ^{n_1+n_2+\cdots+n_N}}{\partial y_1^{n_1} \partial y_2^{n_2} \cdots \partial y_N^{n_N}} \psi(\mathbf{y}) \ \ \  \textrm{for} \ \ \  \mathbf{n} \in \mathbb{Z}^N_{\ge 0},
\end{equation}
while we define the partial antiderivatives, if they exist, as:
\begin{eqnarray}
&&\psi^{(\mathbf{n}-\mathbf{m})}(\mathbf{y})= \psi^{(n_1-m_1,n_2-m_2,\cdots,n_N-m_N)}(\mathbf{y}) \triangleq \partial_{\mathbf{y}}^{\mathbf{n}-\mathbf{m}}\psi(\mathbf{y})\nonumber \\
&=& \underbrace{\int dy_1 \cdots \int dy_1}_{m_1 \ \textrm{times}} 
\cdots
\underbrace{\int dy_N \cdots \int dy_N}_{m_N \ \textrm{times}}
\psi^{(n_1,n_2,\cdots,n_N)}(\mathbf{y})  \\
&& \textrm{for} \ \ \  n_1,\cdots,n_N \in \mathbb{Z}_{\ge 0} \ \ \ \textrm{and} \ \ \ m_1,\cdots,m_N \in \mathbb{Z}_{\ge 0}.
\end{eqnarray}
We use these expressions to express the continuous field and material coefficients, using Taylor expansion polynomials, in the vicinity of a set of discretised points.
It is important to extend the definition of the partial antiderivatives to negative integers with the antiderivatives, even though the analytical expression of the antiderivatives is not always possible, since some governing equations such as viscoelastic wave equations require a memory variable in time, which necessitates the time integral over the past history of the field.

Likewise, due to the local Cartesian assumption~\ref{eq:localCartesian}, we allow us to use the partial derivatives and antiderivatives for a function $\psi(y_{n})$ of one component $y_{n^\prime}$ as:
\begin{equation}
  \psi^{(n)}(y_{n^\prime}) \triangleq \frac{\partial^n}{\partial y_{n^\prime}^n} \psi(y_{n^\prime}) \ \ \ \textrm{and} \ \ \ \psi^{(n-m)}(y_{n^\prime}) \triangleq \int dy_{n^\prime}^m \psi^{(n)}(y_{n^\prime}).
\end{equation}

\subsubsection{Analytical linear partial differential operators}
If the partial differential operator $\mathcal{L}_{ji}$ is quasi-linear, we can write:
\begin{eqnarray}
  \mathcal{L}_{ji} &=& \sum_{\mathbf{n}^\prime \in \mathbb{Z}^N}\sum_{\mathbf{n}\in \mathbb{Z}^N} \left [\partial^{\mathbf{n}^\prime}_{\mathbf{y}}\alpha_{\mathbf{n}^\prime \mathbf{n}ji} \right ] \partial^{\mathbf{n}}_{\mathbf{y}}  \\
  &\triangleq& \sum_{n^\prime_1=-\infty}^\infty \cdots \sum_{n^\prime_N=-\infty}^\infty
  \sum_{n_1=-\infty}^\infty \cdots \sum_{n_N=-\infty}^\infty
  \frac{\partial ^{n^\prime_1+\cdots+n^\prime_N}}{\partial y_1^{n^\prime_1} \cdots \partial y_N^{n^\prime_N}}
  \alpha_{n^\prime_1 \cdots n^\prime_N n_1 \cdots n_N ji}(\mathbf{y})
\frac{\partial ^{n_1+\cdots+n_N}}{\partial y_1^{n_1}  \cdots \partial y_N^{n_N}}, \nonumber 
\end{eqnarray}
where the coefficients $\alpha$ are functions defined in the domain $\Omega$. 
They are related to material properties like density, elastic moduli, viscosity, etc. We call $\alpha$ that are constant as \textit{constant material coefficient} and $\alpha$ that are functions of $\mathbf{y}$ as \textit{variable material coefficient}. 
In some community such as seismology, constant coefficient media are called homogeneous whereas variable coefficient media are heterogeneous. 
We allow the coefficients to be the derivatives and antiderivatives of the material coefficients, since the material coefficients can be coupled to the (anti)derivatives of the field, which can cause a product of the (anti)derivatives of the field and the (anti)derivatives of the material coefficients.
The excitation operator $\mathcal{F}_{jk}$ and the boundary operator $\mathcal{B}_{ji}$ can be written in the same way as $\mathcal{L}_{ji}$:
\begin{equation}
  \mathcal{B}_{ji} = \sum_{\mathbf{n}^\prime\in \mathbb{Z}^N}\sum_{\mathbf{n}\in \mathbb{Z}^N}\left [\partial^{\mathbf{n}^\prime}_{\mathbf{y}}\beta_{\mathbf{n}^\prime\mathbf{n}ji} \right ] \partial^{\mathbf{n}}_{\mathbf{y}},
\end{equation}
and
\begin{equation}
  \mathcal{F}_{jk} = \sum_{\mathbf{n}^\prime\in \mathbb{Z}^N}\sum_{\mathbf{n}\in \mathbb{Z}^N}\left [\partial^{\mathbf{n}^\prime}_{\mathbf{y}}\gamma_{\mathbf{n}^\prime\mathbf{n}jk} \right ] \partial^{\mathbf{n}}_{\mathbf{y}}.
\end{equation}

\subsubsection{Discretisation and Taylor expansion}
We evaluate any function $\psi$, such as the field $u_i$, the source term $f_k$ and the material coefficients $\alpha$, at a discretised nodet:
\begin{equation}
\mathbf{y}_{\pmb{\nu}} = \mathbf{y}_\mathbf{0}+ \sum_{n^\prime=1}^N  \nu_{n^\prime} \Delta y_{n^\prime} \mathbf{e}_{n^\prime}  \ \ \ \textrm{for} \ \ \ \pmb{\nu} \in \mathbf{O} \subset \mathbb{Z}^N,
\end{equation}
where $\mathbf{y}_\mathbf{0}$ the centre (or reference) of the coordinates. $\Delta y_{n^\prime}$ denotes the interval, constant for the simplicity throughout this paper, but it can be variable. $\mathbf{O}$ is the set of indices of the discretised points, with a total number of $|\mathbf{O}|$. For the sake of simplicity, we consider a `Cartesian box' where we have $\mathbf{O} = \{ \nu_{1l}, \nu_{1l}+1, \cdots, \nu_{1r}-1, \nu_{1r} \} \times \cdots \times \{ \nu_{Nl}, \nu_{Nl}+1, \cdots, \nu_{Nr}-1, \nu_{Nr} \} $ and thus $|\mathbf{O}|=(\nu_{1l}-\nu_{1r}+1)\cdots (\nu_{Nr}-\nu_{Nr}+1)$.


In order to simplify the expression of the function and their derivatives and antiderivatives, we use  $[ \ ]$ instead of $( \ )$ to denote the discretised values:
\begin{equation}
\psi[\pmb{\nu}] \triangleq \psi(\mathbf{y}_{\pmb{\nu}}) \ \ \ \textrm{and} \ \ \ \psi^{(\mathbf{l})}[{\pmb{\nu}}] \triangleq \psi^{(\mathbf{l})}(\mathbf{y}_{\pmb{\nu}}) \ \ \ \textrm{for} \ \ \ \mathbf{l} \in \mathbb{Z}^N.
\end{equation}
We then propose the Taylor expansion of the function $\psi$ around the point $\mathbf{y}_{\pmb{\nu}}$:
\begin{eqnarray}
\psi(\mathbf{y}) 
&=& \sum_{l_1=0}^\infty \cdots \sum_{l_N=0}^\infty \frac{\left( y_1 - y_{\nu_1} \right)^{l_1} \cdots \left( y_N - y_{\nu_N} \right)^{l_N}}{l_1!\cdots l_N!} \psi^{(l_1,\cdots,l_N)}[\pmb{\nu}]\nonumber \\
&=& \sum_{\mathbf{l}\in \mathbb{Z}^N_{\ge 0}} \psi^{(\mathbf{l})}[\pmb{\nu}] K_\mathbf{l} (\mathbf{y}-\mathbf{y}_{\pmb{\nu}}) ,
\label{eq:Taylor}
\end{eqnarray}
if we define the kernel function $K_\mathbf{l}$ as:
\begin{equation}
K_\mathbf{l} (\mathbf{y}-\mathbf{y}_{\pmb{\nu}}) \triangleq \frac{\left( y_1 - y_{\nu_1} \right)^{l_1} \left( y_2 - y_{\nu 2} \right)^{l_2} \cdots \left( y_N - y_{\nu_N} \right)^{l_N}}{l_1!l_2!\cdots l_N!}.
\end{equation}
$\psi^{(\mathbf{l})}[\pmb{\nu}]$ being coefficients, we are able to approximate the function $\psi$ in the vicinity of the point $\mathbf{y}_{\pmb{\nu}}$ by using the truncated version of equation~\ref{eq:Taylor}, if needed.

Since this localised polynomial expansion is derivable, we can evaluate the partial derivatives as follows:
\begin{equation}
 \psi^{(n_1, \cdots, n_N)}(\mathbf{y})=\sum_{l_1=n_1}^\infty \cdots \sum_{l_N=n_N}^\infty  \frac{\left( y_1 - y_{\nu_1} \right)^{l_1-n_1}  \cdots \left( y_N - y_{\nu_N} \right)^{l_N-n_N}}{(l_1-n_1)!\cdots(l_N-n_N)!} \psi^{(l_1,\cdots,l_N)}[\pmb{\nu}],
 \label{eq:partialderivativesTaylor}
\end{equation}
with $n_1,\cdots,n_N$ positive integer or zero, for the time being.
However, as equation~\ref{eq:partialderivativesTaylor} is continuous (and thus integrable), we can let $n_1, \cdots, n_N$ to be any integers, including negative values, and we can extend the expression for the mixed derivatives and antiderivatives as follows:
\begin{equation}
  \psi^{(\mathbf{n})}(\mathbf{y}) = \sum_{\mathbf{l}-\mathbf{n}\in \mathbb{Z}_{\ge 0}} \psi^{(\mathbf{l})}[\pmb{\nu}] K_{\mathbf{l}-\mathbf{n}} (\mathbf{y}-\mathbf{y}_{\pmb{\nu}}) ,
  \label{eq:compactTaylor}
\end{equation}
since we are able to propose antiderivative values of the function $\psi$ at the point $\mathbf{y}_{\pmb{\nu}}$ by setting the constant of integration to zero.

\subsubsection{Evaluation of partial derivatives with node values}
Equation~\ref{eq:Taylor} is used to evaluate the partial derivatives of the field and material coefficients at the discretised points.
We first formally describe the values on the nodes around $\mathbf{y}_{\pmb{\nu}}$:
\begin{equation}
  \mathbf{y}_{\pmb{\nu}+\pmb{\eta}} =   \mathbf{y}_{\pmb{\nu}} + \sum_{l=1}^N  \eta_{l} \Delta y_{l} \mathbf{e}_{l}  \ \ \ \textrm{for} \ \ \ \pmb{\eta} \in 
  \{ -L_{l1}, \cdots, L_{r1} \} \times \cdots \times  \{ -L_{lN},\cdots ,L_{rN} \},
\end{equation}
where $L_{l}$ and $L_{r}$ are the left and right numbers of stencils to be taken into account respectively, and they are zero or positive. These values can vary as a function of $\pmb{\nu}$ in order that $\pmb{\nu}+\pmb{\eta}\in \mathbf{O}$. The total number of points to be considered in one dimension minus one, $L_{l1}+L_{r1}$, is \textit{NOT} necessarily equal to the largest degree of partial derivatives that we need to control. We propose the Taylor expansion of the function $\psi$ around the point $\mathbf{y}_{\pmb{\nu}}$ using the truncated version of equation~\ref{eq:Taylor}.
\begin{eqnarray}
  \psi[ {\pmb{\nu}} + {\pmb{\eta}}] &\simeq& \sum_{l_1=0}^{d_1} \cdots \sum_{l_N=0}^{d_N} \frac{\left( \eta_1 \Delta y_1 \right)^{l_1} \cdots \left( \eta_N \Delta y_N \right)^{l_N}}{n_1!\cdots l_N!} \psi^{(l_1,l_2,\cdots,l_N)}[\pmb{\nu}] \nonumber \\
  &=& \sum_{\mathbf{l}\in \mathbf{D}} \psi^{(\mathbf{l})}[\pmb{\nu}] K_{\mathbf{l}} (\pmb{\eta} \odot \Delta \pmb{y})
  \label{eq:truncatedTaylor}
  \end{eqnarray}
  with $\mathbf{D}= \{0,\cdots, d_1 \} \times \cdots \{0,\cdots, d_N\}$ the desired degree of partial derivatives to be controlled and $\odot$ denotes the Hadamard (piecewise) product.
  When the number of elements in $\mathbf{D}$ is larger than the number of points to be considered $(L_{l1}+L_{r1})\times \cdots( L_{lN}+L_{rN})$ this equation is ill-posed, which we use the least-square method to solve the system of equations, by proposing the general inverse of this system of equations.
  By solving the system of equations, we can obtain the values of the partial derivatives of the function $\psi$ at the points $\mathbf{y}_{\pmb{\nu}+{\pmb{\eta}}}$, yielding the approximation of the partials:
  \begin{equation}
    \tilde{\psi}^{(\mathbf{l})}[\pmb{\nu}]= \sum_{\pmb{\eta}} C_{\pmb{\eta}}^{(\mathbf{l})}  \psi[ {\pmb{\nu}} + {\pmb{\eta}}] \ \ \textrm{for} \ \ \mathbf{l} \in \mathbf{D},
    \label{eq:inverseTaylor}
  \end{equation}
where the coefficients $C_{\pmb{\eta}}^{(\mathbf{l})}$ are the elements of the inverse of the matrix of the system of the equations~\ref{eq:truncatedTaylor}, which have only $\Delta \mathbf{y}$ dependence.
Throughout this study, we developed a numerical code using Symbolics.jl \cite{Gowda_2021}, which we will present in the near future. However, due to a slow computation of the inverse of symbolic matrix, these coefficients will be calculated numerically in the real applications, to construct the optimally accurate operators.
If we use directly these results to solve the strong form of the partial differential equation~\ref{eq:strongForm}, truncated only up to the same order as $(L_{l1}+L_{r1})\times \cdots( L_{lN}+L_{rN})$, this method is called a finite-difference method.

Note that we only discuss the collocated stencils in this paper, but the method can be extended to the non-collocated stencils, such as the staggered grid, by extending the expressions above to the different geometry.

\subsection{Weak formulation and test functions}
The strong form of the partial differential equation~\ref{eq:strongForm} is equivalent to the following equation:
\begin{equation}
\forall W \in U, \forall \mathbf{y} \in \Omega \Leftrightarrow W\left [ \mathcal{L}_{ji} u_i (\mathbf{y}) - \mathcal{F}_{jk} f_k (\mathbf{y}) \right ]  =0,
\label{eq:weakForm}
\end{equation} 
and the boundary conditions:
\begin{equation}
  \forall W \in U, \forall \mathbf{y} \in \partial \Omega \Leftrightarrow W\left [ \mathcal{B}_{ji} u_i (\mathbf{y}) \right ]  =0.
\end{equation}
If we propose a bilinear form for a \textit{test} function $W\in U$ as follows:
\begin{equation}
  \int_\Omega \diff[N]{\mathbf{y}} W(\mathbf{y}) \left [ \mathcal{L}_{ji} u_i (\mathbf{y}) - \mathcal{F}_{jk} f_k (\mathbf{y}) \right ] 
  + \int_{\partial \Omega} \diff[N-1]{\mathbf{y}} W(\mathbf{y})\left [ \mathcal{B}_{ji} u_i (\mathbf{y}) \right ]  =0,
  \label{eq:weakFormInt}
\end{equation} 
supposing that the boundary conditions are normalised to have the same dimensions as the volume integrals.
In the discretised form with elements or points that are linked to nodes $\pmb{\nu} \in \mathbf{O}$, the number of unknowns is $M|\mathbf{O}|$, the number of components of $u_i$ times the number of collocated stencils, thus we need to construct a set of at least $|\mathbf{O}|$ test functions applied to approximate the requirement in equation~\ref{eq:weakForm}. 
The choice of the test function set is unlimited but
if we use a set of delta functions collocated on the discretised nodes:
\begin{equation}
W_{\pmb{\nu}} (\mathbf{y}-\mathbf{y}_{\pmb{\nu}})= \delta (\mathbf{y}-\mathbf{y}_{\pmb{\nu}}),
\label{eq:deltaTestFunction}
\end{equation}
the set of $|\mathbf{O}|$ equations~\ref{eq:weakFormInt} is equivalent to the finite-difference method. With this set of delta functions, we do not take into account the physical phenomenon that takes place between the discretised nodes.

\subsection{Review I: Galerkin method -- most of the finite element methods}
In order to take into account the contribution from the area between the nodes, the Galerkin method, the core of the most of the finite-element related discretisation methods \cite{Canuto_2006}, proposes to use a set of locally continuous non-zero functions:
\begin{equation}
  W_{\pmb{\nu}} (\mathbf{y}-\mathbf{y}_{\pmb{\nu}})\ne 0 \ \ \ \textrm{with} \ \ \ \exists R >0, |\mathbf{y}-\mathbf{y}_{\pmb{\nu}}| < R.
\end{equation}
If we can expand the field $u_i$ in terms of a summation of the same basis functions as the set of test functions:
\begin{equation}
  u_i(\mathbf{y}) = \sum_{\pmb{\nu}^\prime} c_{i\pmb{\nu}^\prime} W_{\pmb{\nu}^\prime} (\mathbf{y}-\mathbf{y}_{\pmb{\nu}^\prime}),
  \label{eq:expansion}
\end{equation}
with $c_{i\pmb{\nu}^\prime}$ unknown constant coefficients. Due to the linearity of the operators $\mathcal{L}_{ji},\mathcal{B}_{jk},\mathcal{F}_{ji}$, equation~\ref{eq:weakFormInt} can be rewritten in a matrix form for a set of $W_{\pmb{\nu}}$ test functions, as follows:
\begin{equation}
\mathbf{Gc}=\mathbf{g}
\end{equation}
with the forward modelling operator:
\begin{eqnarray}
  G_{ji\pmb{\nu}\pmb{\nu}^\prime}&=& \int_\Omega \diff[N]{\mathbf{y}} W_{\pmb{\nu}} (\mathbf{y}) \mathcal{L}_{ji} W_{\pmb{\nu}^\prime} (\mathbf{y})  + \int_{\partial \Omega} \diff[N-1]{\mathbf{y}}  W_{\pmb{\nu}} (\mathbf{y}) \mathcal{B}_{ji} W_{\pmb{\nu}^\prime} (\mathbf{y})  \\
  &=& \sum_{\mathbf{n}^\prime \in \mathbb{Z}^N} \sum_{\mathbf{n}\in \mathbb{Z}^N} \left [\int_\Omega \diff[N]{\mathbf{y}} \alpha_{\mathbf{n}^\prime \mathbf{n}ji} W_{\pmb{\nu}}  W_{\pmb{\nu}^\prime}^{(\mathbf{n}^\prime + \mathbf{n})}   + \int_{\partial \Omega} \diff[N-1]{\mathbf{y}} \beta_{\mathbf{n}^\prime \mathbf{n}ji} W_{\pmb{\nu}}  W_{\pmb{\nu}^\prime}^{(
    \mathbf{n}^\prime + \mathbf{n}  )} \right ] \nonumber
\end{eqnarray}
and the external source term:
\begin{equation}
  g_{j\pmb{\nu}} = \int_\Omega \diff[N]{\mathbf{y}} W_{\pmb{\nu}} (\mathbf{y}) \mathcal{F}_{jk} f_k (\mathbf{y}).
\label{eq:sourceGalerkin}
\end{equation}
Locally, $\alpha, \beta, \gamma$ can be nearly constant or extended with a small number of Tayloer expansion terms, which can thus interesting to prepare the set of convolution kernels such as $\int_\Omega \diff[N]{\mathbf{y}}W_{\pmb{\nu}} W^{(\mathbf{n})}_{\pmb{\nu}^\prime}$.
However, the incompleteness of the basis function set $W_{\pmb{\nu}^\prime}$ in equation~\ref{eq:expansion} will lead to numerical errors  that the series of studies on the optimally accurate operators proposed to modify. In fact, all the studies that propose modified operators for specific cases in seismology verify the optimal accuracy based on Taylor expansion, but the derivation of operators has been rather hidden. 
A great success of spectral element methods in seismology is due to the number of freedom that completes the wave propagation phenomenon with high-order Lagrangian polynomials, an aftermath of which is the less sparse matrix system to be solved. 

\subsection{Review II: Distributional finite difference method}
Recently, in seismology the \textit{distributional finite difference method} is introduced \cite[e.g.][]{Masson_2022,Masson_2023,Masson_2024}, that benefits from different sets of (piecewise) test functions to control the accuracy of different degrees of partial derivatives.
Here in this paper, we just show that the distributional finite difference method is one form of finite element methods, based on the weak formulation, and the optimally accurate operators that we will propose in the next section
is related to this method but without a predefined set of basis functions (as $B_{\pmb{\nu}^\prime\mathbf{n}}$ defined below).

If we were able to expand the partial derivatives of $u_i$, $f_k$ and $\alpha$ on different basis functions for each degree of partial derivatives, we could propose the following expansion for the field $u_i$:
\begin{equation}
u_i^{(\mathbf{n})}(\mathbf{y}) = \sum_{\pmb{\nu}^\prime} b_{i\pmb{\nu}^\prime\mathbf{n}} B_{\pmb{\nu}^\prime\mathbf{n}} (\mathbf{y}-\mathbf{y}_{\pmb{\nu}^\prime}),
\label{eq:DFDMexpansion}
\end{equation}
since there is a hope for staggered-grid scheme-like methods, that a set of basis functions $B_{\pmb{\nu}^\prime\mathbf{n}}$ different from $B_{\pmb{\nu}^\prime \mathbf{0}}$ should be constructed with a small number of nodes due to the anti-symmetry of the first-order partial derivatives, 
i.e. the first-order partial derivatives at the middle point can be computed only with two points with the accuracy of $\mathcal{O}(|\Delta \mathbf{y}|^2/4)$, 
while the collocated scheme needs at least three points (by muting the middle point) with the accuracy of $\mathcal{O}(|\Delta \mathbf{y}^2|)$.
Hence, the distributional finite difference method seeks how to map the expansion coefficients $b_{i\pmb{\nu}^\prime \mathbf{n}}$, projected on the specific basis $B_{\pmb{\nu}^\prime\mathbf{n}}$, onto the basis functions $B_{\pmb{\nu}^\prime\mathbf{0}}$, 
especially when those basis functions are \textit{staggered}. 

In order to evaluate the expansion coefficients $b_{i\pmb{\nu}^\prime\mathbf{n}}$ and $b_{i\pmb{\nu}^\prime\mathbf{0}}$, one can evaluate the weighted residuals defined as the difference between the \textit{more accurate} approximation 
$u_i^{(\mathbf{n})}$ (equation~\ref{eq:DFDMexpansion}) 
and the \textit{pragmatic way of expression} $\tilde{u}_i^{(\mathbf{n})}$, which can be derived from equaion~\ref{eq:DFDMexpansion} setting $\mathbf{n}=\mathbf{0}$, as follows:
\begin{eqnarray}
0&\simeq&\int_\Omega \diff[N]{\mathbf{y}} B_{\pmb{\nu}^{\prime \prime} \mathbf{n}} \left ( u_i^{(\mathbf{n})}(\mathbf{y})-\tilde{u}_i^{(\mathbf{n})}(\mathbf{y})\right ) \nonumber \\
&=& \sum_{\pmb{\nu}^{\prime}} \left [  \int_\Omega \diff[N]{\mathbf{y}} B_{\pmb{\nu}^{\prime \prime} \mathbf{n}} b_{i\pmb{\nu}^{\prime} \mathbf{n}} B_{\pmb{\nu}^{\prime}\mathbf{n}} 
-\int_\Omega \diff[N]{\mathbf{y}} B_{\pmb{\nu}^{\prime \prime} \mathbf{n}} b_{i\pmb{\nu}^{\prime }\mathbf{0}} B_{\pmb{\nu}^{\prime }\mathbf{0}}^{(\mathbf{n})}  \right ].
\end{eqnarray}
If we propose a pair of matrices $\mathbf{M}^{(\mathbf{n})}$ and $\mathbf{K}^{(\mathbf{n})}$ as such:
\begin{eqnarray}
  M_{\pmb{\nu}^{\prime \prime}\pmb{\nu}^{\prime}}^{(\mathbf{n})} &=& \int_\Omega \diff[N]{\mathbf{y}} B_{\pmb{\nu}^{\prime \prime}\mathbf{n}} B_{\pmb{\nu}^\prime \mathbf{n}} \nonumber \\
  K_{\pmb{\nu}^{\prime \prime}\pmb{\nu}^{\prime}}^{(\mathbf{n})} &=& \int_\Omega \diff[N]{\mathbf{y}} B_{\pmb{\nu}^{\prime \prime}\mathbf{n}} B_{\pmb{\nu}^\prime \mathbf{0}}^{(\mathbf{n})},
\end{eqnarray}
we can write the relation between the coefficients $b_{i\pmb{\nu}^{\prime}\mathbf{n}}$ and $b_{i\pmb{\nu}^{\prime}\mathbf{0}}$, which we represent $\mathbf{b}_{\mathbf{n}}$ and $\mathbf{b}$ as vectors repsectively, as:
\begin{equation}
\mathbf{b}_{\mathbf{n}} = \mathbf{M}^{(\mathbf{n})-1} \mathbf{K}^{(\mathbf{n})} \mathbf{b}.
\end{equation}
The matrix form of the weak form equation~\ref{eq:weakFormInt} can be rewritten as:
\begin{equation}
\mathbf{Hb}=\mathbf{g}
\end{equation}
with the forward modelling operator:
\begin{eqnarray}
&&H_{ji\pmb{\nu} \pmb{\nu}^\prime} \nonumber \\
&=&  \sum_{\pmb{\nu}^{\prime \prime}}\sum_{\mathbf{n}^\prime \in \mathbb{Z}^N}\sum_{\mathbf{n}\in \mathbb{Z}^N}\left \{  \int_\Omega \diff[N]{\mathbf{y}} \alpha_{\mathbf{n}^\prime \mathbf{n}ji} W_{\pmb{\nu}}  B_{\pmb{\nu}^\prime\mathbf{0}} \left [ \mathbf{M}^{(\mathbf{n}^\prime +\mathbf{n})-1} \mathbf{K}^{(\mathbf{n}^\prime +\mathbf{n})} \right ]_{\pmb{\nu}^{\prime \prime}\pmb{\nu}^{\prime}} \right . \nonumber \\
  &&  \left . \ \ + \int_{\partial \Omega} \diff[N-1]{\mathbf{y}} \beta_{\mathbf{n}^\prime \mathbf{n}ji} W_{\pmb{\nu}}  B_{\pmb{\nu}^\prime\mathbf{0}} \left [ \mathbf{M}^{(\mathbf{n}^\prime +\mathbf{n})-1} \mathbf{K}^{(\mathbf{n}^\prime +\mathbf{n})} \right ]_{\pmb{\nu}^{\prime \prime}\pmb{\nu}^{\prime}}  \right \}
\end{eqnarray}
with the same source term as equation~\ref{eq:sourceGalerkin}.
For variable material coefficients and source terms, we can develop other sets of basis functions to evaluate the matrix elements above but here in this paper, it is not the main focus and the reader can refer to the original paper for more details.
As was also the case with the Galerkin method, the assumption of the completeness of the predefined basis functions (equation~\ref{eq:DFDMexpansion}) seems to be the main limitation of the distributional finite difference method. 
A careful tuning of basis functions (fourth-order B-spline functions in the series of pioneer studies by Y. Masson and his collaborators) is a key to the success and it performs even better than spectral element methods for seismic wave propagation.
Even though theoretically they can extend this method for time-marching scheme, for the moment the application of the distributional finite difference method is limited to the spatial discretisation, using distributional staggered grids.

In this study, however, we want to propose more general framework for 1D to 4D partial differential equations, including time domain, for different types of physics. This is why we propose to use the Taylor expansion instead of
pre-defined basis functions. However, we use the same type of test functions to construct the weak formulations to be solved.

\subsection{Optimally accurate operators: our method}
We \textit{define} here a \textit{new} ``optimally accurate operator'' based on the weak formulation (equation~\ref{eq:weakForm}) of an arbitrary linear(ised) partial differential equation (equation~\ref{eq:strongForm}). 
We use the (locally) continuous non-zero test functions as the Galerkin method, which are associated with the nodes $\pmb{\nu} \in \mathbf{O}$: 
\begin{equation}
  W_{\pmb{\nu}} (\mathbf{y}-\mathbf{y}_{\pmb{\nu}})\ne 0 \ \ \ \textrm{with} \ \ \ \exists R >0, |\mathbf{y}-\mathbf{y}_{\pmb{\nu}}| < R.
  \label{eq:nonZeroTestFunctions}
\end{equation}
However, we expand the field $u_i$, the source term $f_k$ and the material coefficients $\alpha$ with the Taylor expansion that we developed in equation~\ref{eq:compactTaylor}.
We thus have the following expressions for all the ingredients of the governing equation~\ref{eq:strongForm}~and~\ref{eq:boundaryConditionStrongForm}:
\begin{equation}
  \mathcal{L}_{ji}u_i(\mathbf{y}) = \sum_{\mathbf{n}^\prime \in \mathbb{Z}^N} \sum_{\mathbf{n} \in \mathbb{Z}^N} 
  \sum_{\mathbf{l}^\prime-\mathbf{n}^\prime \in \mathbb{Z}^N_{\ge 0}} \sum_{\mathbf{l}-\mathbf{n}\in \mathbb{Z}^N_{\ge 0}}
  \alpha^{(\mathbf{l}^\prime)}_{\mathbf{n}^\prime \mathbf{n}ji}[\pmb{\nu}]  u_i^{(\mathbf{l})}[\pmb{\nu}] K_{\mathbf{l}^\prime-\mathbf{n}^\prime} (\mathbf{y}-\mathbf{y}_{\pmb{\nu}})  K_{\mathbf{l}-\mathbf{n}} (\mathbf{y}-\mathbf{y}_{\pmb{\nu}}) ,
\end{equation}
\begin{equation}
  \mathcal{B}_{ji}u_i(\mathbf{y}) = \sum_{\mathbf{n}^\prime \in \mathbb{Z}^N} \sum_{\mathbf{n} \in \mathbb{Z}^N} 
  \sum_{\mathbf{l}^\prime-\mathbf{n}^\prime \in \mathbb{Z}^N_{\ge 0}} \sum_{\mathbf{l}-\mathbf{n}\in \mathbb{Z}^N_{\ge 0}}
  \beta^{(\mathbf{l}^\prime)}_{\mathbf{n}^\prime \mathbf{n}ji}[\pmb{\nu}]  u_i^{(\mathbf{l})}[\pmb{\nu}] K_{\mathbf{l}^\prime-\mathbf{n}^\prime} (\mathbf{y}-\mathbf{y}_{\pmb{\nu}})  K_{\mathbf{l}-\mathbf{n}} (\mathbf{y}-\mathbf{y}_{\pmb{\nu}}) ,
\end{equation}
\begin{equation}
  \mathcal{F}_{jk}u_i(\mathbf{y}) = \sum_{\mathbf{n}^\prime \in \mathbb{Z}^N} \sum_{\mathbf{n} \in \mathbb{Z}^N} 
  \sum_{\mathbf{l}^\prime-\mathbf{n}^\prime \in \mathbb{Z}^N_{\ge 0}} \sum_{\mathbf{l}-\mathbf{n}\in \mathbb{Z}^N_{\ge 0}}
  \gamma^{(\mathbf{l}^\prime)}_{\mathbf{n}^\prime \mathbf{n}jk}[\pmb{\nu}]  f_k^{(\mathbf{l})}[\pmb{\nu}] K_{\mathbf{l}^\prime-\mathbf{n}^\prime} (\mathbf{y}-\mathbf{y}_{\pmb{\nu}})  K_{\mathbf{l}-\mathbf{n}} (\mathbf{y}-\mathbf{y}_{\pmb{\nu}}) .
\end{equation}
The derivatives on the descritised nodes are not straightforwardly obtained but we can approximate them with equation~\ref{eq:inverseTaylor},
and expand equation~\ref{eq:weakFormInt} with the test functions $W_{\pmb{\nu}}$ as follows:
\begin{eqnarray}
  &&\sum_{\pmb{\eta}^\prime}\sum_{\pmb{\eta}}
  \sum_{\mathbf{n}^\prime \in \mathbb{Z}^N} \sum_{\mathbf{n} \in \mathbb{Z}^N} 
  \sum_{\mathbf{l}^\prime-\mathbf{n}^\prime \in \mathbb{Z}^N_{\ge 0}} \sum_{\mathbf{l}-\mathbf{n}\in \mathbb{Z}^N_{\ge 0}}
 C_{\pmb{\eta}^\prime}^{(\mathbf{l}^\prime)}C_{\pmb{\eta}}^{(\mathbf{l})} \times \nonumber \\
 && \left \{  \alpha_{\mathbf{n}^\prime \mathbf{n}ji}[\pmb{\nu}+\pmb{\eta}^\prime]  u_i[\pmb{\nu}+\pmb{\eta}] 
 - \gamma_{\mathbf{n}^\prime \mathbf{n}jk}[\pmb{\nu}+\pmb{\eta}^\prime]  f_k[\pmb{\nu}+\pmb{\eta}]  \right \} 
  \int_\Omega \diff[N]{\mathbf{y}} W_{\pmb{\nu}}(\mathbf{y}) K_{\mathbf{l}^\prime-\mathbf{n}^\prime} (\mathbf{y}-\mathbf{y}_{\pmb{\nu}})  K_{\mathbf{l}-\mathbf{n}} (\mathbf{y}-\mathbf{y}_{\pmb{\nu}})  \nonumber \\
  &&+  C_{\pmb{\eta}^\prime}^{(\mathbf{l}^\prime)}C_{\pmb{\eta}}^{(\mathbf{l})} \{ \beta_{\mathbf{n}^\prime \mathbf{n}ji}[\pmb{\nu}+\pmb{\eta}^\prime]  u_i[\pmb{\nu}+\pmb{\eta}] \} \int_{\partial \Omega} \diff[N-1]{\mathbf{y}} W_{\pmb{\nu}}(\mathbf{y}) K_{\mathbf{l}^\prime-\mathbf{n}^\prime} (\mathbf{y}-\mathbf{y}_{\pmb{\nu}})  K_{\mathbf{l}-\mathbf{n}} (\mathbf{y}-\mathbf{y}_{\pmb{\nu}})  \nonumber \\
  &\simeq& 0.
  \label{eq:OPTgeneral}
\end{eqnarray}
We can write the matrix form of the optimally accurate operator as:
\begin{equation}
  \mathbf{Av}=\pmb{\Gamma}\mathbf{g}
  \label{eq:OPTmatrix}
\end{equation}
with field and source vectors:
\begin{equation}
v_{i\pmb{\nu}^\prime} = u_i[\pmb{\nu}^\prime]; \ \ g_{k\pmb{\nu}^\prime} = f_k[\pmb{\nu}^\prime],
\end{equation}
and forward modelling operator:
\begin{eqnarray}
&&A_{ji\pmb{\nu} \pmb{\nu}^\prime} \nonumber \\
&=& \sum_{\pmb{\eta}^\prime}
\sum_{\mathbf{n}^\prime \in \mathbb{Z}^N} \sum_{\mathbf{n} \in \mathbb{Z}^N} 
\sum_{\mathbf{l}^\prime-\mathbf{n}^\prime \in \mathbb{Z}^N_{\ge 0}} \sum_{\mathbf{l}-\mathbf{n}\in \mathbb{Z}^N_{\ge 0}} \nonumber \\
&&C_{\pmb{\eta}^\prime}^{(\mathbf{l}^\prime)}C_{\pmb{\nu}^\prime-\pmb{\nu}}^{(\mathbf{l})}  \alpha_{\mathbf{n}^\prime \mathbf{n}ji}[\pmb{\nu}+\pmb{\eta}^\prime]  
\int_\Omega \diff[N]{\mathbf{y}} W_{\pmb{\nu}}(\mathbf{y}) K_{\mathbf{l}^\prime-\mathbf{n}^\prime} (\mathbf{y}-\mathbf{y}_{\pmb{\nu}})  K_{\mathbf{l}-\mathbf{n}} (\mathbf{y}-\mathbf{y}_{\pmb{\nu}})  \nonumber \\
&&+   C_{\pmb{\eta}^\prime}^{(\mathbf{l}^\prime)}C_{\pmb{\nu}^\prime-\pmb{\nu}}^{(\mathbf{l})} \beta_{\mathbf{n}^\prime \mathbf{n}ji}[\pmb{\nu}+\pmb{\eta}^\prime]    \int_{\partial \Omega} \diff[N-1]{\mathbf{y}} W_{\pmb{\nu}}(\mathbf{y}) K_{\mathbf{l}^\prime-\mathbf{n}^\prime} (\mathbf{y}-\mathbf{y}_{\pmb{\nu}})  K_{\mathbf{l}-\mathbf{n}} (\mathbf{y}-\mathbf{y}_{\pmb{\nu}})  \nonumber \\
\end{eqnarray}
with the source operator:
\begin{eqnarray}
 &&\Gamma_{jk\pmb{\nu} \pmb{\nu}^\prime}  \\
  &=& \sum_{\pmb{\eta}^\prime}
  \sum_{\mathbf{n}^\prime \in \mathbb{Z}^N} \sum_{\mathbf{n} \in \mathbb{Z}^N} 
  \sum_{\mathbf{l}^\prime-\mathbf{n}^\prime \in \mathbb{Z}^N_{\ge 0}} \sum_{\mathbf{l}-\mathbf{n}\in \mathbb{Z}^N_{\ge 0}} \nonumber \\
&&    C_{\pmb{\eta}^\prime}^{(\mathbf{l}^\prime)}C_{\pmb{\nu}^\prime-\pmb{\nu}}^{(\mathbf{l})}
    \gamma_{\mathbf{n}^\prime \mathbf{n} jk}[\pmb{\nu}+\pmb{\eta}^\prime]
    \int_\Omega \diff[N]{\mathbf{y}} W_{\pmb{\nu}}(\mathbf{y}) K_{\mathbf{l}^\prime-\mathbf{n}^\prime} (\mathbf{y}-\mathbf{y}_{\pmb{\nu}})  K_{\mathbf{l}-\mathbf{n}} (\mathbf{y}-\mathbf{y}_{\pmb{\nu}}) .\nonumber
\end{eqnarray}
The observation of this equation is that the inverse of Taylor expansion $C_{\pmb{\eta}^\prime}^{(\mathbf{l}^\prime)}$ and $C_{\pmb{\eta}}^{(\mathbf{l})}$ can be prepared in advance symbolically or numerically.
Since the inverse of symbolic matrix is costly for multi dimensions, we derive these coefficients numerically in the real applications.
Material coefficients $\alpha$, boundary conditions $\beta$ and source terms $\gamma$ can be given by an digitalised input model.
The volume integral or surface integral of convolutions between kernel functions and test functions can further be developed by taking benefit of the integral by parts since since the product of kernel functions is a polynomial function, 
which has always an antiderivative. 

If $W_{\pmb{\nu}}$ is differentiable $k_1$ times for $y_1$ and if $\Omega$ is a Cartesian box of $N$ dimension for simplicity, the integral of the product of the kernel functions $F(\mathbf{y})= K_{\mathbf{l}^\prime-\mathbf{n}^\prime} (\mathbf{y}-\mathbf{y}_{\pmb{\nu}})  K_{\mathbf{l}-\mathbf{n}} (\mathbf{y}-\mathbf{y}_{\pmb{\nu}})$ and the test functions can be written as:
\begin{eqnarray}
  && \int_{\Omega} \diff[N]{\mathbf{y}}  W_{\pmb{\nu}}(\mathbf{y}) F(\mathbf{y})\nonumber \\
    &=& \int_{\Omega} \diff[N]{\mathbf{y}} \frac{\partial}{\partial y_1} \left [ W_{\pmb{\nu}} F^{(-1,0,\cdots,0)} \right ] - \int_{\Omega} \diff[N]{\mathbf{y}}  \left [ W_{\pmb{\nu}}^{(1,0,\cdots,0)} F^{(-1,0,\cdots,0)} \right ] \nonumber \\
   &=& \int_{y_1=-\min,\max+\textrm{discon}} \diff[N-1]{\mathbf{y}}  \left [ W_{\pmb{\nu}} F^{(-1,0,\cdots,0)} \right ]  - \int_{\Omega} \diff[N]{\mathbf{y}}  \left [ W_{\pmb{\nu}}^{(1,0,\cdots,0)} F^{(-1,0,\cdots,0)} \right ]  \nonumber \\
   &=& \sum_{i_1=0}^{k_1-1} (-1)^{i_1} \int_{y_1=-\min,\max+\textrm{discon}}  \diff[N-1]{\mathbf{y}} W_{\pmb{\nu}}^{(i_1,0,\cdots,0)} F^{(-i_1-1,0,\cdots, 0)} \nonumber \\
   &&+(-1)^{k_1} \int_\Omega  \diff[N]{\mathbf{y}} W_{\pmb{\nu}}^{(k_1,0,\cdots,0)} F^{(-k_1,0,\cdots,0)}.
 \end{eqnarray}
Note that the above operation used integral by parts.
If $W_{\pmb{\nu}}$ is differentiable $k_1,\cdots,k_N=\mathbf{k}$ times for $\mathbf{y}$ and the $\mathbf{k}$-th derivative is zero everywhere in $\Omega$, we can further develop the above expression as:
\begin{equation}
\int_{\Omega} \diff[N]{\mathbf{y}}  W_{\pmb{\nu}}(\mathbf{y}) F(\mathbf{y})\nonumber \\
  =\sum_{\mathbf{i}=\mathbf{0}}^{\mathbf{k}-\mathbf{1}} (-1)^{\mathbf{i}^T \mathbf{1}} \left .  W_{\pmb{\nu}}^{(\mathbf{i})} F^{(-\mathbf{i}-\mathbf{1})}  \right |_{\mathbf{y}=\textrm{extremities}}.
  \label{eq:generalKernelIntegral}
\end{equation} 
Since the product of kernel functions $F$ is a polynomial function, the antiderivative of $F$ is a polynomial function, which can be computed easily.

We do not develop more in general this expression but we will show several concrete examples in the next section.
Equation~\ref{eq:OPTgeneral} can be solved for $u_i$ iteratively, which does not require a matrix inversion, which we propose in our new method but in order to compare our operators to the \textit{legendary} optimally accurate operators,
we will show the matrix form of the equation~\ref{eq:OPTgeneral} at the same time.

If we use a set of delta functions (equation~\ref{eq:deltaTestFunction}) as the test functions, the volume integral will be zero except for the terms $\mathbf{l}^\prime=\mathbf{n}^\prime $ and $\mathbf{l}=\mathbf{n}$, 
which will lead to the finite-difference method.
As we already discussed earlier, we can propose another set of kernel functions that have a centre in the middle point of the interval between the nodes, in order to mimic the staggered-grid scheme, 
as the distributional finite difference method does. 
However, in this contribution, we will not further discuss this point but the reorganisation of the formulation is straightforward.


In the latter section showing the example for 1D SH wave equation in frequency domain, we will show the equivalence (and a slight difference) between the optimally accurate operators proposed in the series of studies on the simulation of seismic wave propagation and \textit{our new operator}.
Instead of modifying operators based on the Galerkin method, we are litterally dealing with equation~\ref{eq:weakFormInt}.

\section{B-spline basis functions and their properties}
In order to homogeneously take into account the local continuous contribution from the area between nodes, it is common to use the following set of test functions with $\iota$-th order B-spline basis functions:
\begin{equation}
W_{\pmb{\nu}} (\mathbf{y})= \prod_{n=1}^{N} b_{\nu_{n},\iota}(y_{n}) \ \ \ \textrm{for} \ \ \ \mathbf{y} \in \Omega.
\end{equation}
The zero-th order B-spline basis functions are defined as:  
\begin{equation}
b_{\nu_n,0}(y_n) \triangleq \left \{
\begin{array}{ll}
1 & \textrm{if} \  y_{\nu_n} \le y_n  < y_{\nu_n+1 }\\
0 & \textrm{otherwise} 
\end{array}  
\right .\ \ \textrm{for}  \ \ \nu_n \in \{ \nu_{nl}, \nu_{nl}+1, \cdots, \nu_{nr}-1 \}.
\end{equation}
This zero-th order B-spline function set is not useful since the number of members is inferior to the number of node points $|\mathbf{O}|$ for the collocated scheme. 
Note that we truncate the B-spline basis functions at the boundaries of the domain, i.e. $y_{\nu_{nl}}$ and $y_{\nu_{nr}-1}$.
We then define the higher order B-spline basis functions by reccurence:
\begin{equation}
b_{\nu_n,\iota}(y_n) \triangleq \frac{y_n-y_{\nu_n}}{y_{\nu_{n}+\iota}-y_{\nu_n}} b_{\nu_n,\iota-1}(y_n) + \frac{y_{\nu_{n}+\iota+1}-y_n}{y_{\nu_{n}+\iota+1}-y_{\nu_{n+1}}} b_{\nu_{n}+1,\iota-1}(y_n)
\label{eq:BsplineReccurence}
\end{equation}
\begin{equation*}
\ \ \textrm{for} \ \ \nu_n  \in \{ \nu_{nl}- \iota, \nu_{nl}-\iota +1, \cdots, \nu_{nr}-1 \}
\end{equation*}
\begin{equation*}
  \ \ \textrm{and} \ \ y_{\nu_{nl}-i} = y_{\nu_{nl}}-i \Delta y_n, \  y_{\nu_{nr}+i} = y_{\nu_{nr}} +  i \Delta y_n   \ \ \textrm{for} \ \ i \in  \{ 1,\cdots \iota \}.
\end{equation*}
The dummy nodes defined at the last line are required to avoid zero denominators in the reccurence equation since $\Delta y_n$ is positive real.
The properties of these B-spline functions are i) their $(\iota+1)$-times derivability due to their polynomial forms (for a constant interval):
\begin{eqnarray}
  &&b_{\nu_n,\iota}^{(1)}(y_n) 
  = \frac{1}{\iota \Delta y_n} \left [  b_{\nu_n,\iota-1}(y_n) - b_{\nu_{n}+1,\iota-1}(y_n) \right ] \nonumber \\
  && +  \frac{1}{\iota \Delta y_n} \left [(y_n-y_{\nu_n}) b_{\nu_n,\iota-1} ^{(1)}(y_n)
  + (y_{\nu_{n}+\iota+1}-y_n) b_{\nu_{n}+1,\iota-1}^{(1)}(y_n)  \right ]
\end{eqnarray}
and ii) the unicity of their sum on the segment: 
\begin{equation}
\sum_{\nu_n=\nu_{nl}- \iota}^{\nu_{nr}-1} b_{\nu_n,\iota}(y_n) = 1 \ \ \textrm{for} \ \ y_n \in [y_{\nu_{nl}},y_{\nu_{nr}}],
\end{equation}
which assures the same continuous weight on the physical phenomenon throughout the space and time $\Omega$.

In particular, the first-order B-spline function is called a hat-function and they have the same number as that of nodes $|\mathbf{O}|$, thus throughout this paper, we will focus on this specific function set. We have the following properties for the hat-functions:
\begin{equation}
  b_{\nu_n,1}(y_n) =
  \begin{cases}
  \frac{y_n-(\nu_n-1)\Delta y_n}{\Delta y_n} &\textrm{for} \  \ x \in [(\nu_n-1) \Delta y_n , \nu_n\Delta y_n]\cap [y_{\nu_{nl}},y_{\nu_{nr}}] \\
  \frac{-y_n+(\nu_n+1)\Delta y_n}{\Delta y_n} & \textrm{for} \ \ x \in ]\nu_n \Delta y_n , (\nu_n+1)\Delta y_n]\cap [y_{\nu_{nl}},y_{\nu_{nr}}] \\
  0 & \textrm{otherwise} 
  \end{cases},
  \label{eq:testfunctionIn1D}
\end{equation}
with the first derivatives:
\begin{equation}
  b_{\nu_n,1}^{(1)}(y_n) (x)=
  \begin{cases}
  \frac{1}{\Delta y_n} &\textrm{for} \  \ x \in ](\nu_n-1) \Delta y_n , \nu_n \Delta y_n [\cap [y_{\nu_{nl}},y_{\nu_{nr}}] \\
  -\frac{1}{\Delta y_n} &\textrm{for} \ \ x \in ]\nu_n \Delta y_n , (\nu_n+1) \Delta y_n [\cap [y_{\nu_{nl}},y_{\nu_{nr}}] \\
  0 & \textrm{otherwise} 
  \end{cases}
  \label{eq:hatFunction1stDerivative}
\end{equation}
they are discontinuous and we have the second derivatives:
\begin{equation}
b_{\nu_n,1}^{(2)}(y_n) = 0  \ \ \textrm{for} \ \ y_n \in [y_{\nu_{nl}},y_{\nu_{nr}}].
\end{equation}
The higher-order B-spline will be important to enhance the continuity nature of the physics and their function set has more than $|\mathbf{O}|$ elements, thus the method that we want to develop in the near future using them will be an overdetermined problem.

\section{Symbolic example: 1D SH wave equation in frequency domain}
Since the theory above is quite abstract, we will show a concrete example for the 1D SH wave equation in frequency domain, to discuss our proposed method, compared to the original works on the optimally accurate operators, and their relations to other methods.
The specific form of equations~\ref{eq:strongForm}~and~\ref{eq:boundaryConditionStrongForm} are given by setting $M=1$, $N=1$, $y_1=x$, $\Omega=[x_{\nu_l} , x_{\nu_r} ]$ and $\partial \Omega = \{ x_{\nu_l} \} \cup \{x_{\nu_r}\}$, with material coefficients replaced by $\alpha_{00}=\rho \omega^2$, $\alpha_{11}=\mu$, $\alpha_{02}=\mu$, $\beta_{01}=\mu$, $\gamma_{00}=-1$, otherwise zero,  and $\rho$ material density, $\omega$ angular order (constant through $\forall x$), $\mu$ shear modulus, we have:
\begin{equation}
\mathcal{L}= \rho \omega^2 +(\partial_x \mu) \partial_x + \mu \partial_x^2, \ \ \mathcal{B}=\mu \partial_x, \ \ \mathcal{F}=-1,
\end{equation}
which ends up with the following governing equation:
\begin{equation}
  \rho \omega^2 u(x) +  \frac{\partial}{\partial x} \left [ \mu\frac{\partial}{\partial x}u (x) \right ]= -f(x) \ \ \textrm{for} \ \ x \in [x_{\nu_l} , x_{\nu_r}],
  \label{eq:1DSHhomoFreq}
\end{equation}
with a natural (free-surface) boundary condition:
\begin{equation}
  \mu \frac{\partial}{\partial x}u (x)= 0 \ \ \textrm{for} \ \ x \in \{ x_{\nu_l} \} \cup \{x_{\nu_r}\}.
  \label{eq:1DSHhomoFreqBC}
\end{equation}

The modelling matrix $\mathbf{A}$ of equation~\ref{eq:OPTmatrix} can be written as:
\begin{eqnarray}
  A_{\nu \nu^\prime} &=& 
  \sum_{\eta^\prime} \sum_{n^\prime=0,1} \sum_{n=0,1,2} \sum_{l^\prime\ge n^\prime} \sum_{l \ge n} \nonumber \\
 && C_{\eta^\prime}^{({l}^\prime)}C_{\nu^\prime-\nu}^{(l)}
  \alpha_{n^\prime n}[{\nu}+{\eta}^\prime]  
  \int_{x_{\nu l}}^{x_{\nu r}} dx W_{{\nu}}(x) \frac{(x-x_\nu)^{l^\prime-n^\prime + l - n}}{(l^\prime-n^\prime)!(l-n)!}  \\
  && +  C_{{\eta}}^{({l}^\prime)}C_{{\nu^\prime-\nu}}^{({l})} \beta_{{n^\prime n}}[{\nu}+{\eta}^\prime] 
  \left [ W_{{\nu}}(x_{\nu_l}) \frac{(x_{\nu_l}-x_\nu)^{l^\prime-n^\prime + l - n}}{(l^\prime-n^\prime)!(l-n)!} 
  +W_{{\nu}}(x_{\nu_r}) \frac{(x_{\nu_r}-x_\nu)^{l^\prime-n^\prime + l - n}}{(l^\prime-n^\prime)!(l-n)!}  \right ] .\nonumber
  \label{eq:OPTmatrixelement1DseismoFreq}
\end{eqnarray}
Here in 1D case, we can estimate the inverse of Taylor expansion coefficients $C_{\eta}^{(l)}$ symbolically.
If we use three points to compute up to $l-n=4$ or $l^\prime-n^\prime=4$ and set $\eta=\{0,1,2\}$ for $\nu_l<\nu<\nu_r$, we have to inverse this ill-posed matrix:
\begin{equation}
  \mathbf{T}=\left [ 
    \begin{array}{ccccc}
      1 & \Delta x & \frac{1}{2}\Delta x^2 & \frac{1}{6}\Delta x^3 & \frac{1}{24}\Delta x^4 \\
      1 & 0 & 0 & 0 & 0 \\
      1 & -\Delta x & \frac{1}{2}\Delta x^2 & -\frac{1}{6}\Delta x^3 & \frac{1}{24}\Delta x^4 \end{array}
    \right ].
\end{equation}
Here we use the general inverse:
\begin{equation}
\mathbf{T}^\dagger = \mathbf{T}^T(\mathbf{T} \mathbf{T}^T)^{-1} .
\end{equation}
obtaining:
\begin{eqnarray}
  \left [ 
    \begin{array}{ccc}
      C_{-1}^{(0)} & C_{0}^{(0)} & C_{1}^{(0)} \\
      C_{-1}^{(1)} & C_{0}^{(1)} & C_{1}^{(1)} \\
      C_{-1}^{(2)} & C_{0}^{(2)} & C_{1}^{(2)} \\
      C_{-1}^{(3)} & C_{0}^{(3)} & C_{1}^{(3)} \\
      C_{-1}^{(4)} & C_{0}^{(4)} & C_{1}^{(4)} 
    \end{array}
    \right ]
    &=&
    \left [ 
      \begin{array}{ccc}
        0 & 1& 0 \\
        -\frac{1}{2\Delta x +\frac{1}{18}\Delta x^5} & 0 & \frac{1}{2\Delta x +\frac{1}{18}\Delta x^5}  \\
        \frac{1}{\Delta x^2 + \frac{1}{144}\Delta x^6}& -\frac{2}{\Delta x^2 + \frac{1}{144}\Delta x^6} & \frac{1}{\Delta x^2 + \frac{1}{144}\Delta x^6}\\
        - \frac{\Delta x}{12 + \frac{1}{3}\Delta x^4} &0 & \frac{\Delta x}{12 + \frac{1}{3}\Delta x^4} \\
       \frac{1}{12+\frac{1}{12}\Delta x^4} & -\frac{2}{12+\frac{1}{12}\Delta x^4} & \frac{1}{12+\frac{1}{12}\Delta x^4}
      \end{array}
      \right ]  \nonumber \\
    &\simeq&
    \left [ 
      \begin{array}{ccc}
        0 & 1& 0 \\
        -\frac{1}{2\Delta x} & 0 & \frac{1}{2\Delta x} \\
        \frac{1}{\Delta x^2} & -\frac{2}{\Delta x^2} & \frac{1}{\Delta x^2} \\
       - \frac{\Delta x}{12 } &0 & \frac{\Delta x}{12} \\
        \frac{1}{12} & -\frac{1}{6} & \frac{1}{12}
      \end{array}
      \right ] .
      \label{eq:Cleta_approx}
  \end{eqnarray}
We can derive these coefficients for the $\nu=\nu_l$:

Meanwhile, we set the test function as the first-order B-spline basis functions (equation~\ref{eq:testfunctionIn1D}):
\begin{equation}
  W_\nu(x) = b_{\nu,1}(x),
\end{equation}
which is discontinuous only at the first derivation (equation~\ref{eq:hatFunction1stDerivative})
and from the properties of the B-spline basis functions and equation~\ref{eq:generalKernelIntegral}, we have:
\begin{eqnarray}
&&\int_{x_{\nu_l}}^{x_{\nu_r}} dx b_{\nu,1}(x) \frac{(x-x_\nu)^{l^\prime-n^\prime + l - n}}{(l^\prime-n^\prime)!(l-n)!}  \nonumber \\
&=& \left . -b^{(1)}_{\nu,1} (x) \frac{(x-x_\nu)^{l^\prime-n^\prime + l - n+2}}{(l^\prime-n^\prime + l - n+2)(l^\prime-n^\prime + l - n+1)(l^\prime-n^\prime)!(l-n)!}\right |_\textrm{discons}
\nonumber \\
&=&
\begin{cases}
  \frac{\Delta x^{l^\prime-n^\prime + l - n+1}}{(l^\prime-n^\prime + l - n+2)(l^\prime-n^\prime + l - n+1)(l^\prime-n^\prime)!(l-n)!} &\textrm{for} \  \ \nu=\nu_l , \nu_r\\
  \frac{\Delta x^{l^\prime-n^\prime + l - n+1} -(-\Delta x)^{l^\prime-n^\prime + l - n+1}}{(l^\prime-n^\prime + l - n+2)(l^\prime-n^\prime + l - n+1)(l^\prime-n^\prime)!(l-n)!} &\textrm{for} \ \ \nu_l<\nu<\nu_r \\
  \end{cases}
\end{eqnarray}
Taking all the above into account using the approximated version of the inverse of Taylor coefficients (equation~\ref{eq:Cleta_approx}), we can derive an expression for the matrix element $A_{\nu \nu^\prime}$ in equation~\ref{eq:OPTmatrixelement1DseismoFreq} which can be applied to the 1D SH wave equation in frequency domain, for $\nu_l<\nu<\nu_r$ up to $l^\prime \le 4$ and $n \le 4$:
\begin{eqnarray}
  A_{\nu (\nu\pm 1)} &=& 
 \frac{\rho_{\nu \pm 1}\omega^2 {\Delta}x }{12}  \left [ \frac{1}{(1+ \frac{1}{144} {\Delta}x^{4})}\right ] \nonumber \\
&& +
 \frac{\mu_{\nu \pm 1}}{\Delta x} \left [
    \frac{1}{1 + \frac{1}{144} {\Delta}x^{4}} 
  + \frac{ {\Delta}x^{4} }{12(12 + \frac{1}{12} {\Delta}x^{4})} 
  + \frac{{\Delta}x^{4} }{6 \left( \frac{12}{{\Delta}x} + \frac{1}{3} {\Delta}x^{3} \right) \left( 2 {\Delta}x + \frac{1}{18} {\Delta}x^{5} \right)} 
   \right ] \nonumber \\
   &\simeq&
   \frac{\rho_{\nu \pm 1}\omega^2 {\Delta}x }{12} + \frac{\mu_{\nu \pm 1}}{\Delta x}
\label{eq:seismo1DSHOPT}
\end{eqnarray}
\begin{eqnarray}
  A_{\nu \nu} &=&
 \rho_{\nu}\omega^2{\Delta}x \left [ 1- \frac{1}{6(1+ \frac{1}{144} {\Delta}x^{4})} \right ] 
 \nonumber \\
  &&
  -\frac{ 2 \mu_\nu  }{{\Delta}x} \left [ \frac{1}{ (1+ \frac{1}{144} {\Delta}x^{4})} 
  + \frac{ {\Delta}x^{4} }{12(12 + \frac{1}{12} {\Delta}x^{4})} 
   \right ] \nonumber \\
   &\simeq&
    \frac{5}{6}\rho_{\nu}\omega^2{\Delta}x - \frac{ 2 \mu_\nu  }{{\Delta}x}
  \end{eqnarray}

As we show in \ref{app1}, this operator without the higher order terms is the same as the one obtained in GT95.

\section{Numerical example: 1D Poisson equation}
As it appears in many physics, we here focus on the 1D Poisson equation to test our method numerically, which is indeed the $\mu$ dependency of the operator for 1D SH wave propagation (equation~\ref{eq:seismo1DSHOPT}). However, this time, we solve these numerically without the truncation of higher orders. If we take the final operator in equation~\ref{eq:seismo1DSHOPT}, it is identical to that of the 3 point convetional finite-difference operator. As is described above, we can no more obtain symbolic expression, which could look like an iteration of fractions to a high order, if we do not truncate.
In order to evaluate the errors to the analytical solution, we set the equations~\ref{eq:strongForm} by $M=1$, $N=1$, $y_1=x$, $\Omega=[0,10\pi ]$ , with material coefficients replaced by  $\alpha_{11}=\kappa$, $\alpha_{02}=\kappa$, $\gamma_{00}=-1$, otherwise zero,  and $\kappa$ conductivity, the field $u_1=T$ the temperature,
\begin{equation}
\partial_x(\kappa (x)\partial_x T(x)) - f(x)  =0.
\label{eq:1Dpoisson}
\end{equation}

\begin{figure}[h]
  \includegraphics[width=0.7\textwidth]{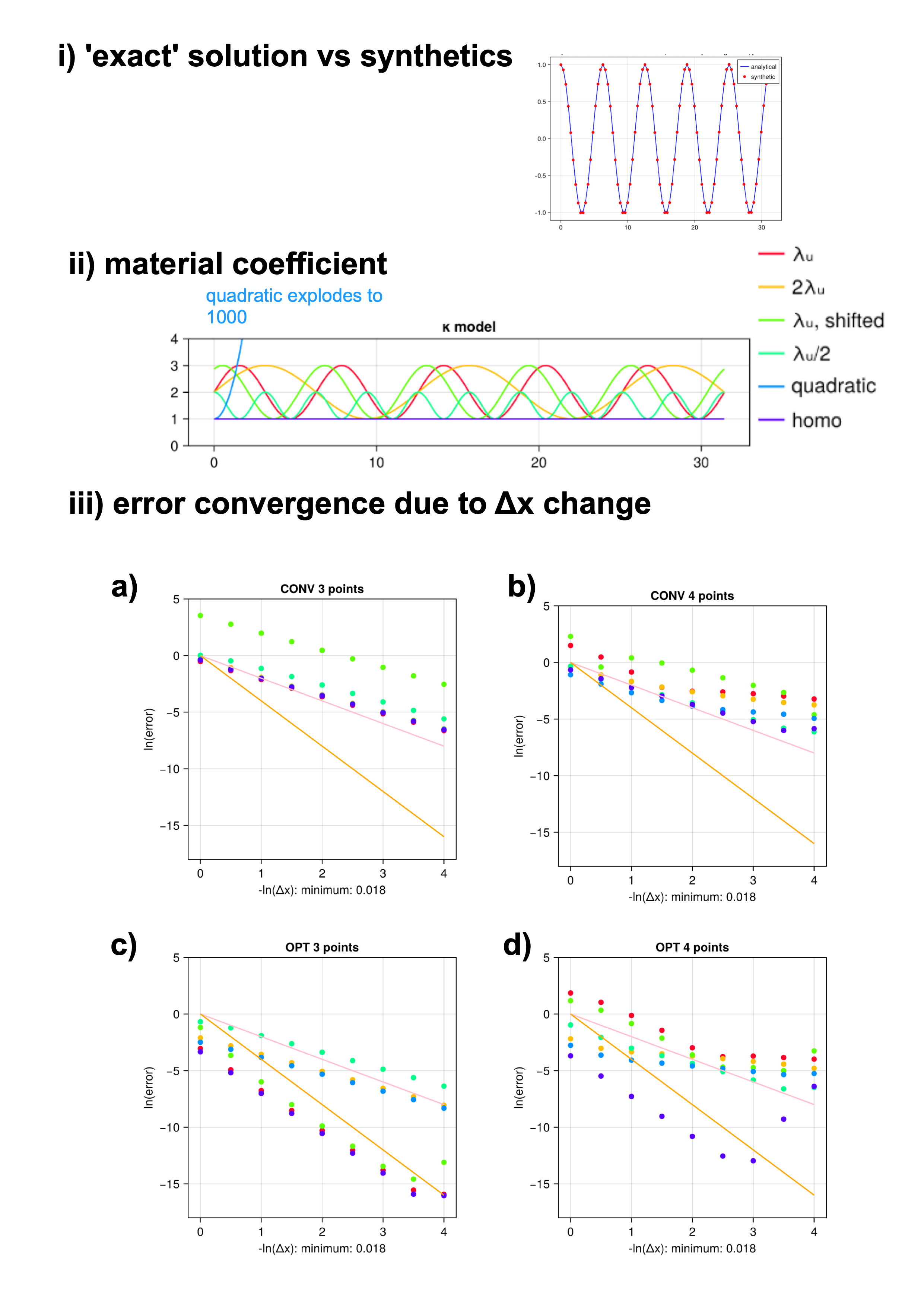}
  \caption{A benchmark test for our new optimally accurate operators: i) the cosine function of $T(x)$ to be obtained; 
  ii) the material coefficient $\kappa$ is defined on the same \mbox{1D} space with different wavelength of sine waves and also quadratic and homogeneous configurations; 
  iii) the convergence test varying $\Delta x = \exp(-t)$ for $t=0,1,\cdots 8$ for 
  a) 3-point conventional finite-difference method,
   b) 4-point conventional finite-difference method, 
   c) 3-point optimally accurate operator with linear B-spline test functions, 
   d) 4-point optimally accurate operator with linear B-spline test functions.}
  \label{fig:convergence1D}
 \end{figure}

Here we propose an exact solution as in Figure~\ref{fig:convergence1D}i):
\begin{equation}
T(x) = \cos(x) \ \ \mathrm{for} \ \ x \in [0,10\pi]
\end{equation}
with explicit boundary conditions:
\begin{equation}
T(0) = 0, \ \ T(10\pi) = 0.
\end{equation}
We give several different material-coefficient scenarii as in Figure~\ref{fig:convergence1D}ii): 
``$\lambda_u$'' for a sine function with the same wavelength $\kappa(x)=\sin(x)+2$; 
``$2\lambda_u$'' for a sine function with a half the wavelength $\kappa(x)=\sin(x/2)+2$, 
``$\lambda_u$ shifted'' for a sine function with the same wavelength with a phase shift $\kappa(x)=\sin(x+\pi/3)+2$,
``$\lambda_u/2$'' for a cosine function with a half the wavelength $\kappa(x)=\cos(x)^2+1$,
``quadratic'' for an explosing second-order polynomial $\kappa(x)=x^2+1$,
``homo'' for a constant $\kappa(x)=1$.
We analytically compute equation~\ref{eq:1Dpoisson} to obtain an expression for $f(x)$. 

We then numerically compute the matrix-vector products in the equation~\ref{eq:OPTmatrix}, using inverse of Taylor expansions with 3 and 4 points.
For the finite-difference scheme, we used $\delta-$functions as test functions.
We follow the Newton scheme as in (https://github.com/tduretz/ExactFieldSolutions.jl) 
using sparse matrix differentiation scheme (https://github.com/JuliaDiff/SparseDiffTools.jl).

We then benchmarked our methodology for various $\Delta x= \exp(-t)$ with various $t=0,1,\cdots 8$ to compute the 
numerical errors for different configurations as in Figure~\ref{fig:convergence1D}iii).

The first observation is that the optimally accurate operator gives $\exp(3-4)$ higher precision for three-points operators.
However, the conventional finite-difference operator can obtain better precision with the four-points operator as expected,
with the linear B-spine test functions, which guarantee the continuity of only three points in the vicinity, 
do not improve the error - indeed, it worsens the precision. 

We also see that our operators have O4 convergence for homogeneous material coefficient and those with the same wavelength 
(``$\lambda_u$',``$\lambda_u$ shifted'), the others show O2 convergence. 

\section{Discussion and Conclusion}
In this paper, we were able to decipher the theory behind optimally accurate operators.
As discussed more explicitly in the following appendix, the \textit{classical} theory from GT95
truncated too early in the evaluation of the bilinear form involving the product between the test function and the Taylor-expanded field.
We demonstrated the similarity between the operators derived from our new theory and those of GT95,
and we performed a numerical benchmark against conventional finite-difference schemes.
Since the GT95 operator for this term is rather “conventional,” as shown in the previous section,
GT95 (and all subsequent studies based on homogeneous media) using simple symbolic expressions
will not be able to reproduce our results.

As this study is still in the exploratory phase, the Newton solver has not yet been optimized.
We will publish our Julia code once the solver is accelerated.
The next step is to assess higher-order B-spline functions to determine whether error convergence
can be improved in all cases.
Another important topic we have not discussed in detail is the treatment of boundary conditions.

\section{Acknowledgements}

This project marks the first phase of the MuseSeLFiE project led by N. Fuji, funded by the Institut Universitaire de France (IUF),
aiming to develop a flexible community code.
During the summer workshop CLEEDI 2023 in Foix, T. Duretz encouraged the use of Julia programming, prompting NF to start working with it.
When NF sought to automate the generation of numerical operators, the idea for the present theory emerged.
The project progressed rapidly during two $\partial$GPU4GEO hackathons held in the Black Forest.

\appendix
\section{Intuitive derivation of the optimally accurate operators and its relation to the original paper of GT95}
\label{app1}

Here in this appendix, we show an intuitive derivation of the optimally accurate operators for the 1D homogeneous SH wave equation in the frequency domain and 2D laplacian operator. Since GT95 did not mention the derivation of the optimally accurate operators, we provide a simple derivation of the optimally accurate operators for the 1D homogeneous SH wave equation in the frequency domain and 2D laplacian operator. Note that the notations are not the same as that of the main text, for the sake of simplicity for those who are familiar with these problems.

\subsection{Hat-function based 1D homogeneous SH elastodynamic equation in frequency domain}
Let us begin with the equation of motion of 1D homogeneous SH wave propagation in frequency domain:
\begin{equation}
  \rho \omega^2 u(x) + \mu \frac{\partial^2}{\partial x^2}u (x)= -f(x) \ \ \textrm{for} \ \ x \in [0,L],
  \label{eq:1DSHhomoFreq}
\end{equation}
with a natural (free-surface) boundary condition:
\begin{equation}
  \mu \frac{\partial}{\partial x}u (x)= 0 \ \ \textrm{for} \ \ x \in \{0\} \cup \{ L \},
  \label{eq:1DSHhomoFreqBC}
\end{equation}
where $\rho$ is the density, $\mu$ is the rigidity, $u$ is the displacement and $f$ the force. 
If we use a set of linear spline test functions $\{ \phi_m(x) \}$ for $m=0,1, \cdots, M$:
\begin{equation}
  \phi_m(x) =
  \begin{cases}
  \frac{x-(m-1)\Delta x}{\Delta x} &\textrm{for} \  \ x \in [(m-1) \Delta x , m\Delta x ]\cap [0,L] \\
  \frac{-x+(m+1)\Delta x}{\Delta x} & \textrm{for} \ \ x \in ]m \Delta x , (m+1)\Delta x]\cap [0,L] \\
  0 & \textrm{otherwise} 
  \end{cases}
\end{equation}
with $\Delta x = L/M$.
This set of test functions is continous and differentiable. The first derivatives are:
\begin{equation}
  \phi_m^{(1)}(x)=
  \begin{cases}
  \frac{1}{\Delta x} &\textrm{for} \  \ x \in ](m-1) \Delta x , m\Delta x [\cap [0,L] \\
  -\frac{1}{\Delta x} & \textrm{for} \ \ x \in ]m \Delta x , (m+1)\Delta x[\cap [0,L] \\
  0 & \textrm{otherwise} 
  \end{cases}
\end{equation}
thus discontinuous 
and the second derivatives are:
\begin{equation}
  \phi_m^{(2)}(x)=0 \ \ \textrm{for} \ \ x \in [0,L].
\end{equation}
Note that the superscript in the parenthesis denotes the number of partial derivativation with respect to $x$ throughout this section.
The choice of the test functions is per se arbitrary but in order to maintain the consinuity of the trial function (the partial diffeerential equation to be solved),
we need to weight equally throughout the domain. B-spline functions have a property that the sum of the functions is 1 anywhere in the domain defined.

\subsubsection{Galerkin method}
The Galerkin method expands the field $u$ in terms of summation of the basis functions $\phi_m$ (the same set as that of the test functions):
\begin{equation}
  u(x)=\sum_{n=0}^M c_n \phi_n(x).
\end{equation}
Since the operator is linear, taking the line integral of the equation of motion (equation~\ref{eq:1DSHhomoFreq}) and the boundary condition (equation~\ref{eq:1DSHhomoFreqBC}), multiplied by the test (and basis) function $\phi_m$ over the domain $[0,L]$, substituiting $u$ by the expansion above, we obtain the following linear system:
\begin{eqnarray}
  &&\sum_{n=0}^M c_n \left [ \rho \omega^2 \int_0^L dx \phi_m(x) \phi_n(x) + \mu \int_0^L dx \phi_m(x) \phi^{(2)}_n(x) -\phi_m (L)\mu \phi^{(1)}_n(L)+\phi_m (0)\mu \phi^{(1)}_n(0)\right ] \nonumber \\
  &=& \sum_{n=0}^M c_n \left [ \rho \omega^2 \int_0^L dx \phi_m(x) \phi_n(x) - \mu \int_0^L dx \phi_m^{(1)} \phi^{(1)}_n (x)\right ] \nonumber \\
  &=& \int_0^L dx \phi_m(x) f(x).
\end{eqnarray}
If we write this in matrix form:
\begin{equation}
  \mathbf{G} \mathbf{c} = -\mathbf{g},
\end{equation}
with
\begin{eqnarray}
  G_{00}=G_{MM} &=& \rho \omega^2 \frac{ \Delta x}{3} - \frac{\mu}{\Delta x}\\
  G_{01}=G_{M(M-1)} &=&  \rho \omega^2 \frac{\Delta x}{6} + \frac{\mu}{\Delta x} \\ 
  G_{mm}&=&  \rho \omega^2 \frac{2 \Delta x}{3} - \frac{2\mu}{\Delta x}\\
  G_{m(m\pm 1)} &=& \rho \omega^2 \frac{\Delta x}{6} + \frac{\mu}{\Delta x} \\
  G_{mn} &=& 0 \ \ \textrm{otherwise}.
  \label{eq:1DSHhomoFreq_GalerkinOperator}
\end{eqnarray}
We can observe the same operator for $\mu u^{(2)}$ in $G_{m(m\pm 1)}$ and $G_{mm}$ as that of the second-order three-point finite difference operator.

\subsubsection{`Generalised' way of optimal operator derivation}
We are interested in finding a linear operator $\mathbf{A}$ that can approximate:
\begin{equation}
  \sum_n A_{mn} v_n \sim \int_0^L dx \phi_m(x) \left [ \rho \omega^2 u(x) + \mu \frac{\partial^2}{\partial x^2}u (x) \right ] + \left [ \phi_m(0) \mu \frac{\partial}{\partial x}u (0) \right ]+ \left [ \phi_m(L) \mu \frac{\partial}{\partial x}u (L) \right ]
  \label{eq:1DSHhomoFreq_aim}
\end{equation}
with a discreted set of displacement:
\begin{equation}
  v_n=u(n\Delta x)
\end{equation}
and the excitation term:
\begin{equation}
  g_m= -\int_0^L dx \phi_m (x) f(x)
\end{equation}
in order to construct a linear system:
\begin{equation}
  \mathbf{A} \mathbf{v} = \mathbf{g}.
\end{equation}

Suppose that $u$ is continuous on $[0,L]$.
Since the $\phi_m(x)$ is localised around $x=m\Delta x$ it is naural to propose a Taylor expansion of $u$ in the vicinity of $x=m\Delta x$ as follows:
\begin{equation}
  u(x)= \sum_{l=0}^\infty \frac{1}{l!} (x-m \Delta x)^l u^{(l)} (m\Delta x) 
  \label{eq:taylorlocal}
\end{equation}
where the superscript in the parenthesis denotes the number of partial derivativation with respect to $x$:
\begin{equation}
  u^{(l)} (x) = \frac{\partial^l}{\partial x^l}u (x).
\end{equation}
For the sake of convenience we denote the partial derivatives $u^{(l)}$ on the nodes as:
\begin{equation}
  v^{(l)}_m=u^{(l)}(m\Delta x)
\end{equation}
This localised polynomial (equation~\ref{eq:taylorlocal}) is  derivable:
\begin{equation}
  u^{(l)}(x) = \sum_{l^\prime=l}^\infty \frac{1}{(l^\prime-l)!} (x-m \Delta x)^{l^\prime-l} u^{(l^\prime)} (m\Delta x) 
\end{equation}
and continous (i.e. integrable) with a set of antiderivatives (with the constant to be zero):
\begin{equation}
  u^{(-l)}(x) = \sum_{l^\prime=\min(0,l)}^\infty \frac{1}{(l^\prime+l)!} (x-m \Delta x)^{l^\prime+l} u^{(l^\prime)} (m\Delta x).
\end{equation}

We evaluate equation~\ref{eq:1DSHhomoFreq_aim}.
\begin{eqnarray}
  &&\int_0^L dx \phi_m(x) \left \{ \textrm{l.h.s. of eq. of motion} \right \} \nonumber \\
  &=& \left . \phi_m(x) \left \{ \textrm{l.h.s. of eq. of motion} \right \}^{(-1)} \right |^L_0
  - \left . \phi_m^{(1)} \left \{ \textrm{l.h.s. of eq. of motion} \right \}^{(-2)} \right |_\textrm{discons}
  \nonumber \\
  &=&
  \begin{cases}
   \displaystyle \dfrac{1}{\Delta x} \left [\rho \omega^2 \sum_{l=0}^\infty \dfrac{\Delta x^{l+2}} {(l+2)!}v^{(l)}_0
     +  \mu \sum_{l=2}^\infty \dfrac{\Delta x^l}{l!} v^{(l)}_0 \right ]&\textrm{for} \ \ m=0 \\ \\
    \displaystyle \frac{2}{\Delta x} \left [ \rho \omega^2 \sum_{l^\prime = 0}^\infty \frac{\Delta x^{2l^\prime+2}}{(2l^\prime+2)!} v^{(2l^\prime)}_m + \mu \sum_{l^\prime=1}^\infty \frac{\Delta x^{2l^\prime}}{(2l^\prime)!} v^{(2l^\prime)}_m \right ]&\textrm{for}\ \ m=1,\cdots, M-1 \\ \\
    \displaystyle \dfrac{1}{\Delta x} \left [\rho \omega^2 \sum_{l=0}^\infty \dfrac{\Delta x^{l+2}} {(l+2)!}v^{(l)}_M
      +  \mu \sum_{l=2}^\infty \dfrac{\Delta x^l}{l!} v^{(l)}_M \right ]&\textrm{for} \ \ m=M \\
  \end{cases} 
  \label{eq:lhs_of_eq_of_motion_1DSHhomoFreq}
\end{eqnarray}
The left hand side of the equation of motion (denoted as ``l.h.s. of eq. of motion'') is the integrand of the first term of the right-hand side of equation~\ref{eq:1DSHhomoFreq_aim} and the first development is obtained by integrating by parts, knowing that $\phi_m^{(0)}$ is continuous and $\phi_m^{(2)}$ is zero everywhere. Taking the boundary conditions into account, the following should be satisfied:
\begin{eqnarray}
  && \sum_n A_{mn} v_n \nonumber \\
  &\simeq&
  \begin{cases}
    \displaystyle \rho \omega^2 \left [  \frac{\Delta x}{2} v^{(0)}_0+ \frac{\Delta x^2}{6}v^{(1)}_0 + \frac{\Delta x^3}{24} v^{(2)}_0 \right ] +\mu 
    \left [ v^{(1)}_0+  \frac{\Delta x}{2} v^{(2)}_0+ \frac{\Delta x^2}{6}v^{(3)}_0 + \frac{\Delta x^3}{24} v^{(4)}_0 \right ] , \  m=0 \\
    \displaystyle \rho \omega^2 \left [  \Delta x v^{(0)}_m + \frac{\Delta x^3}{12}v^{(2)}_m + \frac{\Delta x^5}{360} v^{(4)}_m \right ] 
    + \mu\left [  \Delta x v^{(2)}_m + \frac{\Delta x^3}{12}v^{(4)}_m + \frac{\Delta x^5}{360} v^{(6)}_m \right ] ,\  m=1,\cdots, M-1 \\
    \displaystyle \rho \omega^2 \left [  \frac{\Delta x}{2} v^{(0)}_M+ \frac{\Delta x^2}{6}v^{(1)}_M + \frac{\Delta x^3}{24} v^{(2)}_M \right ] 
    +\mu 
    \left [ v^{(1)}_M+  \frac{\Delta x}{2} v^{(2)}_M+ \frac{\Delta x^2}{6}v^{(3)}_M + \frac{\Delta x^3}{24} v^{(4)}_M \right ] ,\  m=M \\
  \end{cases} \nonumber \\
  &\simeq&
  \begin{cases}
    \displaystyle \rho \omega^2   \frac{\Delta x}{2} v^{(0)}_0 
    +\left (\rho \omega^2 \frac{\Delta x^2}{6}+ \mu \right )v^{(1)}_0  
    + \left ( \rho \omega^2 \frac{\Delta x^3}{24} + \mu \frac{\Delta x}{2}\right )v^{(2)}_0 + \cdots, \  m=0 \\
    \displaystyle \rho \omega^2 \Delta x v^{(0)}_m + \left ( \rho \omega^2 \frac{\Delta x^3}{12} + \mu \Delta x \right ) v^{(2)}_m +
    \left ( \rho \omega^2 \frac{\Delta x^5}{360} + \mu \frac{\Delta x^3}{12} \right ) v^{(4)}_m +
    \cdots, \  m=1,\cdots, M-1 \\
    \displaystyle \rho \omega^2   \frac{\Delta x}{2} v^{(0)}_M 
    +\left (\rho \omega^2 \frac{\Delta x^2}{6}+ \mu \right )v^{(1)}_M  
    + \left ( \rho \omega^2 \frac{\Delta x^3}{24} + \mu \frac{\Delta x}{2}\right )v^{(2)}_M + \cdots, \  m=M \\
  \end{cases}
  \label{eq:algebra_for_1DSHhomoFreq}
\end{eqnarray}
We then use the Taylor expansion for the displacement $v_{m\pm 1}$:
\begin{equation}
  v_{m\pm 1} = v_m \pm \Delta x v^{(1)}_m + \frac{\Delta x^2}{2} v^{(2)}_m \pm \frac{\Delta x^3}{6} v^{(3)}_m + \cdots
\end{equation}
and thus we have:
\begin{eqnarray}
  &&\sum_n A_{mn} v_n \nonumber \\
  &=& \sum_{l=0}^\infty \frac{v_m^{(l)}}{l!} \sum_{n=m-1}^{m+1} A_{mn} \left (n-m \right )^l \Delta x^l\nonumber \\
  &=& \frac{v_m^{(0)}}{0!} \left [ A_{m(m-1)}+ A_{mm} + A_{m(m+1)} \right ]  \nonumber \\
  &+& \frac{v_m^{(1)}}{1!} \left [ -A_{m(m-1)}+ A_{m(m+1)} \right ] \Delta x \nonumber \\
  &+& \frac{v_m^{(2)}}{2!} \left [ A_{m(m-1)}+A_{m(m+1)} \right ] \Delta x^2 + \cdots .
  \label{eq:Av_summation3points}
\end{eqnarray}
By solving equaion~\ref{eq:algebra_for_1DSHhomoFreq} for $v_m^{(0-2)}$, we obtain the following linear operator:
\begin{eqnarray}
  A_{00}=A_{MM} &=& \rho \omega^2 \frac{ \Delta x}{3} - \frac{\mu}{\Delta x}\nonumber \\
  A_{01}=A_{M(M-1)} &=&  \rho \omega^2 \frac{\Delta x}{6} + \frac{\mu}{\Delta x}\nonumber  \\ 
  A_{mm}&=&  \rho \omega^2 \frac{5 \Delta x}{6} - \frac{2\mu}{\Delta x}\nonumber \\
  A_{m(m\pm 1)} &=& \rho \omega^2 \frac{\Delta x}{12} + \frac{\mu}{\Delta x}\nonumber \\
  A_{mn} &=& 0 \ \ \textrm{otherwise}.
  \label{eq:1DSHhomoFreq_operator}
\end{eqnarray}
If one does not require the matrix to be tridiagonal, the boundary operator can have three non-zero elements as well:
\begin{eqnarray}
  A_{00}=A_{MM} &=& \rho \omega^2 \frac{ 19 \Delta x}{24} - \frac{\mu}{\Delta x}\nonumber \\
  A_{01}=A_{M(M-1)} &=&  \rho \omega^2 \frac{\Delta x}{4} + \frac{\mu}{\Delta x}\nonumber \\
  A_{02}=A_{M(M-2)} &=& - \rho \omega^2 \frac{\Delta x}{24}.
\end{eqnarray}
\textcolor{black}{Stepping further, we just need to store the expression (equation~\ref{eq:algebra_for_1DSHhomoFreq}) up to $v_m^{(0-2)}$, substituiting them by the Taylor expansion of $v_{m\pm 1}$ and solving the linear system for $v_{m\pm 1}$ without constructing $A$ explicitly.}

\subsubsection{Comparison with GT95 operators}
The obtained operator for a SH wave propagation in a homogeneous media in frequency domain (equation~\ref{eq:1DSHhomoFreq_operator}) is the same for $A_{mn}$ in the middle with the modified operator proposed by GT95 for the 1D SH wave propagation in frequency domain, except for the operator on the boundaries.
This is due to the evaluation of the integral of equation~\ref{eq:lhs_of_eq_of_motion_1DSHhomoFreq} in GT95 and this study. Here we show that the evaluation of the integral in GT95 is less accurate, which results in this small discrepancy.

Instead of evaluating the continuity of $u$ along $x$, GT95 approximated the integrand in eq. (3.13) by its value at each node $v_n$ to evaluate the integral of the left hand side of equation~\ref{eq:lhs_of_eq_of_motion_1DSHhomoFreq}. 
This is equivalent to make a somehow truncated zeroth order assumption for $i$th partial derivative:
\begin{equation}
  u^{(i)}(x)=
    v_n^{(i)}\ \ \textrm{for} \ \ x \in [\max(0,(n-1)\Delta x), \min(L, (n+1)\Delta x)].
\end{equation}
Hence we have:
\begin{equation}
  \left \{ \rho \omega^2 u + \mu u^{(2)}\right \}^{(-2)}=
    \displaystyle \rho \omega^2\frac{v_n^{(0)} x^2}{2} + \mu \frac{v_n^{(2)}x^2}{2} \ \ \textrm{for} \ \ x \in [\max(0,(n-1)\Delta x), \min(L, (n+1)\Delta x)]
\end{equation}
which leads to:
\begin{eqnarray}
  &&\int_0^L dx \phi_m(x) \left \{ \textrm{l.h.s. of eq. of motion} \right \} + \left [ \phi_m(0) \mu \frac{\partial}{\partial x}u (0) \right ]+ \left [ \phi_m(L) \mu \frac{\partial}{\partial x}u (L) \right ]\nonumber \\
  &=& 
  - \left . \phi_m^{(1)} \left \{ \textrm{l.h.s. of eq. of motion} \right \}^{(-2)} \right |^{x=\min(L, (m+1)\Delta x)}_{x=\max(0,(m-1)\Delta x)} + \left [ \phi_m(0) \mu \frac{\partial}{\partial x}u (0) \right ]+ \left [ \phi_m(L) \mu \frac{\partial}{\partial x}u (L) \right ]\nonumber \\
  \nonumber \\
  &\simeq&
  \begin{cases}
    \displaystyle \rho \omega^2   \frac{\Delta x}{2} v^{(0)}_0 + \mu v^{(1)}_0  + \mu \frac{\Delta x}{2}v^{(2)}_0
    ,\  m=0 \\
    \displaystyle \rho \omega^2 \Delta x v^{(0)}_m  + \mu \Delta x  v^{(2)}_m 
    ,\  m=1,\cdots, M-1 \\
    \displaystyle \rho \omega^2   \frac{\Delta x}{2} v^{(0)}_M + \mu v^{(1)}_M  +\mu \frac{\Delta x}{2}v^{(2)}_M
    , \  m=M \\
  \end{cases}
  \label{eq:lhs_of_eq_of_motion_1DSHhomoFreqGT95}
\end{eqnarray}
This `desired exact value' should be then compared to equation~\ref{eq:Av_summation3points}. At first sight, the comparison up to the second order $v_m^{(0-2)}$ will give us the following operator:
\begin{eqnarray}
    A^{unmod}_{00}=A_{MM} &=& \rho \omega^2 \frac{ \Delta x}{2} - \frac{\mu}{\Delta x}\nonumber \\
    A^{unmod}_{01}=A_{M(M-1)} &=&    \frac{\mu}{\Delta x} \nonumber \\ 
    A^{unmod}_{mm}&=&  \rho \omega^2 \Delta x - \frac{2\mu}{\Delta x}\nonumber \\
    A^{unmod}_{m(m\pm 1)} &=&  \frac{\mu}{\Delta x} \nonumber \\
    A^{unmod}_{mn} &=& 0 \ \ \textrm{otherwise}.
    \label{eq:1DSHhomoFreq_operatorGTunmodified}
\end{eqnarray}
However, the terms related to $\mu$ (stiffness matrix) should generate the higher order terms:
\begin{eqnarray}
 \sum_n A^{unmod}_{mn} v_n &=&
 \begin{cases}
  \displaystyle \rho \omega^2   \frac{\Delta x}{2} v^{(0)}_0 + \mu v^{(1)}_0  + \mu \frac{\Delta x}{2}v^{(2)}_0
  + \mu \frac{\Delta x^2}{6} v^{(3)}_0 + \mu \frac{\Delta x^3}{24} v^{(4)}_0 + \cdots
  ,\  m=0 \\
  \displaystyle \rho \omega^2 \Delta x v^{(0)}_m  + \mu \Delta x  v^{(2)}_m + \mu \frac{\Delta x^3}{12} v^{(4)}_m + \cdots
  ,\  m=1,\cdots, M-1 \\
  \displaystyle \rho \omega^2   \frac{\Delta x}{2} v^{(0)}_M + \mu v^{(1)}_M  +\mu \frac{\Delta x}{2}v^{(2)}_M
  + \mu \frac{\Delta x^2}{6} v^{(3)}_M + \mu \frac{\Delta x^3}{24} v^{(4)}_M + \cdots
  , \  m=M \\
\end{cases}.
\label{eq:Avcontribution_unmodifiedGT95}
\end{eqnarray}
GT95 then propose to use a physical property (`normal mode of the system'):
\begin{equation}
  \rho \omega u^{(2)}(x) + \mu u^{(4)}(x) =0 \ \ \textrm{for} \ \ x \in [0,L].
  \label{eq:1DSHhomoFreq_normalmode}
\end{equation}
We reorganise `desired exact value' by adding this property:
\begin{eqnarray}
  &&\int_0^L dx \phi_m(x) \left \{ \textrm{l.h.s. of eq. of motion} \right \} + \left [ \phi_m(0) \mu \frac{\partial}{\partial x}u (0) \right ]+ \left [ \phi_m(L) \mu \frac{\partial}{\partial x}u (L) \right ]\nonumber \\
  &\simeq&
  \begin{cases}
    \displaystyle \rho \omega^2   \frac{\Delta x}{2} v^{(0)}_0 + \mu v^{(1)}_0  + \mu \frac{\Delta x}{2}v^{(2)}_0
    + \rho \omega^2 \frac{\Delta x^3}{24} v^{(2)}_0 + \mu \frac{\Delta x^3}{24} v^{(4)}_0 
    ,\  m=0 \\
    \displaystyle \rho \omega^2 \Delta x v^{(0)}_m  + \mu \Delta x  v^{(2)}_m 
    + \rho \omega^2 \frac{\Delta x^3}{12} v^{(2)}_m + \mu \frac{\Delta x^3}{12} v^{(4)}_m 
    ,\  m=1,\cdots, M-1 \\
    \displaystyle \rho \omega^2   \frac{\Delta x}{2} v^{(0)}_M + \mu v^{(1)}_M  +\mu \frac{\Delta x}{2}v^{(2)}_M
    + \rho \omega^2 \frac{\Delta x^3}{24} v^{(2)}_M + \mu \frac{\Delta x^3}{24} v^{(4)}_M 
    , \  m=M \\
  \end{cases}
  \label{eq:lhs_of_eq_of_motion_1DSHhomoFreqGT95modified}
\end{eqnarray}
\begin{eqnarray}
  A_{00}^{GT95} =A_{MM}^{GT95}&=& \rho \omega^2 \frac{ 5\Delta x}{12} - \frac{\mu}{\Delta x}\nonumber \\
  A_{01}^{GT95} =A_{M(M-1)}^{GT95}&=&  \rho \omega^2 \frac{\Delta x}{12} + \frac{\mu}{\Delta x} \nonumber  \\ 
  A_{mm}^{GT95}&=&  \rho \omega^2 \frac{5 \Delta x}{6} - \frac{2\mu}{\Delta x}\nonumber \\
  A_{m(m\pm 1)}^{GT95} &=& \rho \omega^2 \frac{\Delta x}{12} + \frac{\mu}{\Delta x}\nonumber \\
  A_{mn}^{GT95} &=& 0 \ \ \textrm{otherwise}.
  \label{eq:1DSHhomoFreq_operatorGT95}
\end{eqnarray}
No explicit derivation of this operator is given in the original GT95 paper, but this is how we can `modify' it if one follows the logic of GT95. However, the `physical' requirements (equation~\ref{eq:1DSHhomoFreq_normalmode}) are not trivial to apply for any other type of physics (viscoelasticity, gravity, etc.). This is one of the reason why the optimally accurate operators have not been widely used.

Except for $A_{00}, A_{01},A_{MM}, A_{M(M-1)}$, GT95 operators are identical to our proposed operator (equation~\ref{eq:1DSHhomoFreq_operator}) by coincidence.
The ignorance of the variation of the field variable and its derivatives along the space in GT95 requires less accurate desired value for $\sum_n A_{mn}u_n$ than the one in this study.

\bibliographystyle{elsarticle-num} 
\bibliography{./bibliography.bib}

\begin{thebibliography}{10}
\expandafter\ifx\csname url\endcsname\relax
  \def\url#1{\texttt{#1}}\fi
\expandafter\ifx\csname urlprefix\endcsname\relax\def\urlprefix{URL }\fi
\expandafter\ifx\csname href\endcsname\relax
  \def\href#1#2{#2} \def\path#1{#1}\fi

\bibitem{Geller_1995}
R.~Geller, N.~Takeuchi, A new method for computing highly accurate {DSM} synthetic seismograms, Geophysical Journal International (1995).
\newblock \href {https://doi.org/10.1111/j.1365-246X.1995.tb06865.x} {\path{doi:10.1111/j.1365-246X.1995.tb06865.x}}.

\bibitem{Geller_1998}
R.~Geller, N.~Takeuchi, Optimally accurate second-order time-domain finite difference scheme for the elastic equation of motion: one-dimensional cases, Geophysical Journal International (1998).

\bibitem{Takeuchi_2000}
N.~Takeuchi, R.~Geller, P.~R. Cummins, Physics of the Earth and Planetary Interiors 119 (2000) 25--36.

\bibitem{Geller_2011}
R.~Geller, H.~Mizutani, N.~Hirabayashi, Existence of a second island of stability of predictor-corrector schemes for calculating synthetic seismograms, Geophysical Journal International 188 (2011) 253--262.

\bibitem{Hasegawa_2016}
K.~Hasegawa, R.~J. Geller, N.~Hirabayashi, \href{https://doi.org/10.1093/gji/ggw079}{An error analysis of higher-order finite-element methods: effect of degenerate coupling on simulation of elastic wave propagation}, Geophysical Journal International 205~(3) (2016) 1532--1547.
\newblock \href {http://arxiv.org/abs/https://academic.oup.com/gji/article-pdf/205/3/1532/17369866/ggw079.pdf} {\path{arXiv:https://academic.oup.com/gji/article-pdf/205/3/1532/17369866/ggw079.pdf}}, \href {https://doi.org/10.1093/gji/ggw079} {\path{doi:10.1093/gji/ggw079}}.
\newline\urlprefix\url{https://doi.org/10.1093/gji/ggw079}

\bibitem{Hasegawa_2018}
K.~Hasegawa, N.~Fuji, K.~Konishi, Improvement of accuracy of the spectral element method for elastic wave computation using modified numerical integration operators, Comp. Methods Appl. Mech. Eng. 342 (2018) 200--223.

\bibitem{Geller_1994}
R.~Geller, T.~Ohminato, Computation of synthetic seismograms and their partial derivatives for heterogeneous media with arbitrary natural boundary conditions using the direct solution~…, Geophys. J. Int. 116 (1994) 421--446.

\bibitem{Cummins_Computation_1997}
P.~Cummins, N.~Takeuchi, R.~Geller, Computation of complete synthetic seismograms for laterally heterogeneous models using the direct solution method, Geophys J Int 130~(1) (1997) 1--16.
\newblock \href {https://doi.org/10.1111/j.1365-246X.1997.tb00983.x} {\path{doi:10.1111/j.1365-246X.1997.tb00983.x}}.

\bibitem{Kawai_Complete_2006}
K.~Kawai, N.~Takeuchi, R.~J. Geller, Complete synthetic seismograms up to 2 hz for transversely isotropic spherically symmetric media, Geophys J Int 164~(2) (2006) 411--424.
\newblock \href {https://doi.org/10.1111/j.1365-246X.2005.02829.x} {\path{doi:10.1111/j.1365-246X.2005.02829.x}}.

\bibitem{Fuji_2012}
N.~Fuji, S.~Chevrot, L.~Zhao, R.~J. Geller, K.~Kawai, Finite-frequency structural sensitivities of short-period compressional body waves, Geophysical Journal International 190~(1) (2012) 522--540.

\bibitem{Fuji_2010}
N.~Fuji, K.~Kawai, R.~J. Geller, A methodology for inversion of broadband seismic waveforms for elastic and anelastic structure and its application to the mantle transition zone beneath the northwestern pacific, Physics of the Earth and Planetary Interiors 180~(3-4) (2010) 118--137.

\bibitem{Konishi_2014}
K.~Kensuke, K.~Kenji, G.~Robert~J., F.~Nobuaki, Geophysical Journal International 199 (2014) 1245--1267.

\bibitem{Konishi_2020}
K.~Konishi, N.~Fuji, F.~Deschamps, Three-dimensional elastic and anelastic structure of the lowermost mantle beneath the western pacific from finite-frequency tomography, Journal of Geophysical Research: Solid Earth 125~(2) (2020) e2019JB018089.

\bibitem{Fuji_2020}
N.~Fuji, Composition of the symphonic poem ses, insight no.1, philogaïa orchestra – from ipgp orchestra to philogaïa orchestra, CIG newsletter (2020).
\newblock \href {https://doi.org/geodynamics.org/cig/index.php?cID=1311)} {\path{doi:geodynamics.org/cig/index.php?cID=1311)}}.

\bibitem{Franken_2020}
T.~Franken, J.~Armitage, N.~Fuji, A.~Fournier, Seismic wave-based constraints on geodynamical processes: An application to partial melting beneath the réunion island, Geochemistry, Geophysics, Geosystems 21~(5) (2020) e2019GC008815.

\bibitem{Jacob_2022}
A.~Jacob, M.~Plasman, C.~Perrin, N.~Fuji, P.~Lognonné, Z.~Xu, M.~Drilleau, N.~Brinkman, S.~Stähler, G.~Sainton, A.~Lucas, D.~Giardini, T.~Kawamura, J.~Clinton, W.~Banerdt, \href{https://www.sciencedirect.com/science/article/pii/S0040195122002281}{Seismic sources of insight marsquakes and seismotectonic context of elysium planitia, mars}, Tectonophysics 837 (2022) 229434.
\newblock \href {https://doi.org/https://doi.org/10.1016/j.tecto.2022.229434} {\path{doi:https://doi.org/10.1016/j.tecto.2022.229434}}.
\newline\urlprefix\url{https://www.sciencedirect.com/science/article/pii/S0040195122002281}

\bibitem{Komatisch_1998}
D.~Komatitsch, J.~Vilotte, The spectral element method: an efficient tool to simulate the seismic response of 2d and 3d geological structures, Bull. Seis. Soc. Am. 88 (1998) 368--392.

\bibitem{Komatitsch_2002}
D.~Komatitsch, J.~Tromp, Spectral-element simulations of global seismic wave propagation---i. validation, Geophysical Journal International 149~(2) (2002) 390--412.

\bibitem{NissenMeyer_2014}
T.~Nissen-Meyer, M.~van Driel, S.~C. St\"ahler, K.~Hosseini, S.~Hempel, L.~Auer, A.~Colombi, A.~Fournier, \href{https://se.copernicus.org/articles/5/425/2014/}{Axisem: broadband 3-d seismic wavefields in axisymmetric media}, Solid Earth 5~(1) (2014) 425--445.
\newblock \href {https://doi.org/10.5194/se-5-425-2014} {\path{doi:10.5194/se-5-425-2014}}.
\newline\urlprefix\url{https://se.copernicus.org/articles/5/425/2014/}

\bibitem{Moczo_2007}
P.~Moczo, J.~O. Robertsson, L.~Eisner, \href{https://www.sciencedirect.com/science/article/pii/S0065268706480080}{The finite-difference time-domain method for modeling of seismic wave propagation} 48 (2007) 421--516.
\newblock \href {https://doi.org/https://doi.org/10.1016/S0065-2687(06)48008-0} {\path{doi:https://doi.org/10.1016/S0065-2687(06)48008-0}}.
\newline\urlprefix\url{https://www.sciencedirect.com/science/article/pii/S0065268706480080}

\bibitem{Igel_2016}
H.~Igel, \href{https://doi.org/10.1093/acprof:oso/9780198717409.001.0001}{Computational Seismology: A Practical Introduction}, Oxford University Press, 2016.
\newblock \href {https://doi.org/10.1093/acprof:oso/9780198717409.001.0001} {\path{doi:10.1093/acprof:oso/9780198717409.001.0001}}.
\newline\urlprefix\url{https://doi.org/10.1093/acprof:oso/9780198717409.001.0001}

\bibitem{Lele1992Nov}
S.~K. Lele, {Compact finite difference schemes with spectral-like resolution}, J. Comput. Phys. 103~(1) (1992) 16--42.
\newblock \href {https://doi.org/10.1016/0021-9991(92)90324-R} {\path{doi:10.1016/0021-9991(92)90324-R}}.

\bibitem{Spotz1995Oct}
W.~F. Spotz, G.~F. Carey, {High-order compact scheme for the steady stream-function vorticity equations}, Int. J. Numer. Methods Eng. 38~(20) (1995) 3497--3512.
\newblock \href {https://doi.org/10.1002/nme.1620382008} {\path{doi:10.1002/nme.1620382008}}.

\bibitem{Spotz1996Mar}
W.~F. Spotz, G.~F. Carey, {A high-order compact formulation for the 3D Poisson equation}, Numer. Methods Partial Differ. Equations 12~(2) (1996) 235--243.
\newblock \href {https://doi.org/10.1002/(SICI)1098-2426(199603)12:2<235::AID-NUM6>3.0.CO;2-R} {\path{doi:10.1002/(SICI)1098-2426(199603)12:2<235::AID-NUM6>3.0.CO;2-R}}.

\bibitem{Carey1997Jul}
G.~F. Carey, W.~F. Spotz, {Higher-order compact mixed methods}, Commun. Numer. Methods Eng. 13~(7) (1997) 553--564.
\newblock \href {https://doi.org/10.1002/(SICI)1099-0887(199707)13:7<553::AID-CNM80>3.0.CO;2-O} {\path{doi:10.1002/(SICI)1099-0887(199707)13:7<553::AID-CNM80>3.0.CO;2-O}}.

\bibitem{Abide2020Dec}
S.~Abide, {Finite Difference preconditioning for compact scheme discretizations of the Poisson equation with variable coefficients}, J. Comput. Appl. Math. 379 (2020) 112872.
\newblock \href {https://doi.org/10.1016/j.cam.2020.112872} {\path{doi:10.1016/j.cam.2020.112872}}.

\bibitem{Masson_2022}
Y.~Masson, \href{https://doi.org/10.1093/gji/ggac306}{Distributional finite-difference modelling of seismic waves}, Geophysical Journal International 233~(1) (2022) 264--296.
\newblock \href {http://arxiv.org/abs/https://academic.oup.com/gji/article-pdf/233/1/264/48352323/ggac306.pdf} {\path{arXiv:https://academic.oup.com/gji/article-pdf/233/1/264/48352323/ggac306.pdf}}, \href {https://doi.org/10.1093/gji/ggac306} {\path{doi:10.1093/gji/ggac306}}.
\newline\urlprefix\url{https://doi.org/10.1093/gji/ggac306}

\bibitem{Masson_2023}
Y.~Masson, J.~Virieux, P-sv wave propagation in heterogeneous media: Velocity-stress distributional finite-difference method, Geophysics 1 (2023) 1--92.
\newblock \href {https://doi.org/10.1190/geo2022-0118.1} {\path{doi:10.1190/geo2022-0118.1}}.

\bibitem{Gowda_2021}
S.~Gowda, Y.~Ma, A.~Cheli, M.~Gwozdz, V.~B. Shah, A.~Edelman, C.~Rackauckas, High-performance symbolic-numerics via multiple dispatch, arXiv preprint arXiv:2105.03949 (2021).

\bibitem{Canuto_2006}
C.~Canuto, M.~Y. Hussaini, A.~Quarteroni, T.~A. Zang, Spectral Methods. Fundamentals in Single Domains, Springer-Verlag, Berlin, 2006.

\bibitem{Masson_2024}
Y.~Masson, C.~Lyu, P.~Moczo, Y.~Capdeville, B.~Romanowicz, J.~Virieux, \href{https://doi.org/10.1093/gji/ggae025}{2-d seismic wave propagation using the distributional finite-difference method: further developments and potential for global seismology}, Geophysical Journal International 237~(1) (2024) 339--363.
\newblock \href {http://arxiv.org/abs/https://academic.oup.com/gji/article-pdf/237/1/339/56706602/ggae025.pdf} {\path{arXiv:https://academic.oup.com/gji/article-pdf/237/1/339/56706602/ggae025.pdf}}, \href {https://doi.org/10.1093/gji/ggae025} {\path{doi:10.1093/gji/ggae025}}.
\newline\urlprefix\url{https://doi.org/10.1093/gji/ggae025}

\end{thebibliography}


\begin{thebibliography}{00}


\bibitem{lamport94}
  Leslie Lamport,
  \textit{\LaTeX: a document preparation system},
  Addison Wesley, Massachusetts,
  2nd edition,
  1994.

\end{thebibliography}
\end{document}